\documentclass[aps,amsmath,amssymb,preprintnumbers,groupedaddress,twocolumn]{revtex4}
\usepackage{graphicx,psfrag,hyperref}
\usepackage{amsmath,bbm,array,amsfonts,graphicx,wrapfig,lscape,float,mathtools,multirow,longtable}
\usepackage[dvipsnames]{xcolor}

\newcommand{\be}{\begin{equation}}
\newcommand{\ee}{\end{equation}}
\newcommand{\beq}{\begin{equation}}
\newcommand{\beql}[1]{\begin{equation}\label{#1}}
\newcommand{\eeq}{\end{equation}}
\newcommand{\ba}{\begin{array}}
\newcommand{\ea}{\end{array}}
\newcommand{\bea}{\begin{eqnarray}}
\newcommand{\beal}[1]{\begin{eqnarray}\label{#1}}
\newcommand{\eea}{\end{eqnarray}}
\newcommand{\ben}{\begin{enumerate}}
\newcommand{\een}{\end{enumerate}}
\newcommand{\bean}{\begin{eqnarray*}}
\newcommand{\eean}{\end{eqnarray*}}
\newcommand{\eref}[1]{(\ref{#1})}
\newcommand{\sref}[1]{\S\ref{#1}}
\newcommand{\tref}[1]{Table~\ref{#1}}
\newcommand{\nn}{\nonumber}

\newcommand{\fref}[1]{Figure \ref{#1}}
\newcommand{\btab}[1]{\begin{tabular}{#1}}
\newcommand{\etab}{\end{tabular}}

\def \Black {\pdfsetcolor{0 0 0 1}}

\newcommand{\comment}[1]{}

\newcommand{\ud}{\mathrm{d}}

\newcommand{\qed}{\nobreak \ifvmode \relax \else
      \ifdim\lastskip<1.5em \hskip-\lastskip
      \hskip1.5em plus0em minus0.5em \fi \nobreak
      \vrule height0.75em width0.5em depth0.25em\fi}

\definecolor{darkspringgreen}{rgb}{0.09, 0.45, 0.27}
\definecolor{forestgreen}{rgb}{0.13, 0.55, 0.13}

\begin{document}

\preprint{UNIST-MTH-23-RS-05}
\title{Machine Learning Regularization \\ for the Minimum Volume Formula of Toric Calabi-Yau 3-folds}

\author{Eugene Choi${}^{a}$} 
\author{Rak-Kyeong Seong${}^{a,b}$}
\email{xeugenechoi@gmail.com, seong@unist.ac.kr}

\affiliation{\it ${}^{a}$ 
Department of Mathematical Sciences, and \\ 
 ${}^{b}$ 
Department of Physics,\\ 
Ulsan National Institute of Science and Technology,\\
50 UNIST-gil, Ulsan 44919, South Korea
}

\begin{abstract}
We present a collection of explicit formulas for the minimum volume of Sasaki-Einstein 5-manifolds. 
The cone over these 5-manifolds is a toric Calabi-Yau 3-fold.
These toric Calabi-Yau 3-folds are associated with an infinite class of $4d$ $\mathcal{N}=1$ supersymmetric gauge theories, which are realized as worldvolume theories of D3-branes probing the toric Calabi-Yau 3-folds.
Under the AdS/CFT correspondence, the minimum volume of the Sasaki-Einstein base is inversely proportional to the central charge of the corresponding $4d$ $\mathcal{N}=1$ superconformal field theories.
The presented formulas for the minimum volume are in terms of geometric invariants of the toric Calabi-Yau 3-folds.
These explicit results are derived by implementing machine learning regularization techniques that advance beyond previous applications of machine learning for determining the minimum volume.
Moreover, the use of machine learning regularization allows us to present interpretable and explainable formulas for the minimum volume.
Our work confirms that, even for extensive sets of toric Calabi-Yau 3-folds, the proposed formulas approximate the minimum volume with remarkable accuracy.
\end{abstract} 
\maketitle
\noindent

\section{Introduction}

Since the introduction of machine learning techniques in \cite{He:2017aed,Krefl:2017yox,Ruehle:2017mzq,Carifio:2017bov,Cole:2019enn,Cole:2020gkd,Halverson:2020trp,Gukov:2020qaj,Abel:2021rrj,Krippendorf:2021uxu,Cole:2021nnt,Berglund:2023ztk,Demirtas:2023fir} for studying problems that occur in the context of string theory, 
machine learning -- both supervised \cite{Bull:2018uow,Jejjala:2019kio,Brodie:2019dfx,He:2020lbz,Erbin:2020tks,Anagiannis:2021cco,Larfors:2022nep} and unsupervised \cite{Krippendorf:2020gny,Berman:2021mcw,Bao:2021olg,Seong:2023njx} -- has led to a variety of applications in string theory.
A problem that appeared particularly suited for machine learning in 2017 \cite{Krefl:2017yox} was the problem of identifying a formula for the minimum volume of Sasaki-Einstein 5-manifolds \cite{Martelli:2006yb,Martelli:2005tp}. 
The cone over these Sasaki-Einstein 5-manifolds is a toric Calabi-Yau 3-fold \cite{fulton,1997hep.th...11013L}.
Given that there are infinitely many toric Calabi-Yau 3-folds with corresponding Sasaki-Einstein 5-manifolds 
and that there is an infinite class of $4d$ $\mathcal{N}=1$ supersymmetric gauge theories associated to them via string theory \cite{Greene:1996cy,Douglas:1997de,Witten:1998qj,Klebanov:1998hh, Douglas:1996sw,Lawrence:1998ja,Feng:2000mi,Feng:2001xr}, 
this beautiful correspondence between geometry and gauge theory
was identified in \cite{Krefl:2017yox} as an ideal testbed for introducing machine learning for string theory.

These $4d$ $\mathcal{N}=1$ supersymmetric gauge theories corresponding to toric Calabi-Yau 3-folds are realized as worldvolume theories of D3-branes probing the Calabi-Yau singularities.
Via the AdS/CFT correspondence \cite{Maldacena:1997re,Morrison:1998cs,Acharya:1998db}, the minimum volume of the Sasaki-Einstein 5-manifolds is related to the maximized $a$-function \cite{Intriligator:2003jj,Butti:2005vn,Butti:2005ps} that gives the central charges of the corresponding $4d$ $\mathcal{N}=1$ superconformal field theories \cite{Gubser:1998vd,Henningson:1998gx}. 
The proposal in \cite{Krefl:2017yox} was that machine learning techniques can be used to give a formula of the minimum volume in terms of features taken from the toric diagram of the corresponding toric Calabi-Yau 3-folds.
Such a formula would significantly simplify the computation of the minimum volume, which conventionally is computed by minimizing the volume function obtained from the equivariant index \cite{Martelli:2006yb,Martelli:2005tp} or Hilbert series of the toric Calabi-Yau 3-fold \cite{Benvenuti:2006qr,Feng:2007ur}.

In \cite{Krefl:2017yox}, we made use of multiple linear regression \cite{gauss1823theoria, fisher1922mathematical,mendenhall2003second,  freedman2009statistical,jobson2012applied} and a combination of a regression model and a convolutional neural network (CNN) \cite{lecun1998gradient,krizhevsky2012imagenet,lecun2015deep,schmidhuber2015deep} to learn the minimum volume for toric Calabi-Yau 3-folds. 
As it is often the case for supervised machine learning \cite{rumelhart1986learning,hastie2009elements}, 
the models lacked interpretability and explainability, achieving high accuracies in estimating the minimum volume with giving only little insight into the mathematical structure and physical origin of the estimating formula. 

\begin{table}[h!]
\centering
\begin{tabular}{|c|cccc|cccc|cc|}
\hline
\; & 0 & 1 & 2 & 3 & 4 & 5 & 6 & 7 & 8 & 9 \\
\hline \hline
D5 & $\times$ & $\times$ & $\times$ & $\times$ & $\cdot$ & $\times$ & $\cdot$ & $\times$ & $\cdot$ & $\cdot$
\\
NS5 & $\times$ & $\times$ & $\times$ & $\times$ & \multicolumn{4}{c|}{--- $\Sigma$ ---} & $\cdot$ & $\cdot$
\\
\hline
\end{tabular}
\caption{Type IIB brane configuration for brane tilings, where $\Sigma : P(x,y)=0$ refers to the holomorphic curve defined by the corresponding toric Calabi-Yau 3-fold and the Newton polynomial $P(x,y)$ of the associated toric diagram $\Delta$ \cite{Hori:2000kt,Feng:2005gw}.}
\label{t_brane}
\end{table}

In this work, we aim to highlight the pivotal role of regularization techniques in machine learning \cite{tikhonov1963regularization,hastie2009elements}.
We demonstrate that employing regularized machine learning models can effectively address the limitations inherent in supervised machine learning, especially for problems that appear in string theory and, more broadly, for problems at the intersection of mathematics and physics.
While the primary objective of regularization in machine learning is to prevent overfitting, certain versions of it can be employed to eliminate model parameters, echoing the spirit of regularization in quantum field theory.

By focusing on Least Absolute Shrinkage and Selection Operator (Lasso) regularization \cite{tibshirani1996regression} for polynomial and logarithmic regression models, we identify several candidate formulas for the minimum volume of Sasaki-Einstein 5-manifolds corresponding to toric Calabi-Yau 3-folds.
The discovered formulas depend either on 3 or 6 parameters that come from features of the corresponding toric diagrams \cite{fulton,1997hep.th...11013L} -- convex lattice polygons on $\mathbb{Z}^2$ that characterize uniquely the associated toric Calabi-Yau 3-fold.
Compared to the extremely large number of parameters in the regression and CNN models used in our previous work in \cite{Krefl:2017yox},
the formulas obtained in this study are both presentable, interpretable, and most importantly reusable for the computation of the minimum volume for toric Calabi-Yau 3-folds.

\section{Calabi-Yau 3-Folds and Quiver Gauge Theories}\label{sback2}

In this work, we concentrate on non-compact toric Calabi-Yau 3-folds $\mathcal{X}$.
These geometries can be considered as cones over Sasaki-Einstein 5-manifolds $Y_5$ \cite{Maldacena:1997re,Morrison:1998cs,Acharya:1998db,Martelli:2004wu,Benvenuti:2004dy,Benvenuti:2005ja,Butti:2005sw}.
The toric Calabi-Yau 3-folds are fully characterized by convex lattice polygons $\Delta$ on $\mathbb{Z}^2$ known as toric diagrams \cite{fulton,1997hep.th...11013L}.
The associated Calabi-Yau singularities can be probed by D3-branes whose worldvolume theories form a class of $4d$ $\mathcal{N}=1$ supersymmetric gauge theories \cite{Greene:1996cy,Douglas:1997de,Witten:1998qj,Klebanov:1998hh, Douglas:1996sw,Lawrence:1998ja,Feng:2000mi,Feng:2001xr}.

\begin{figure}[ht!!]
\begin{center}
\resizebox{0.85\hsize}{!}{
\includegraphics[height=5cm]{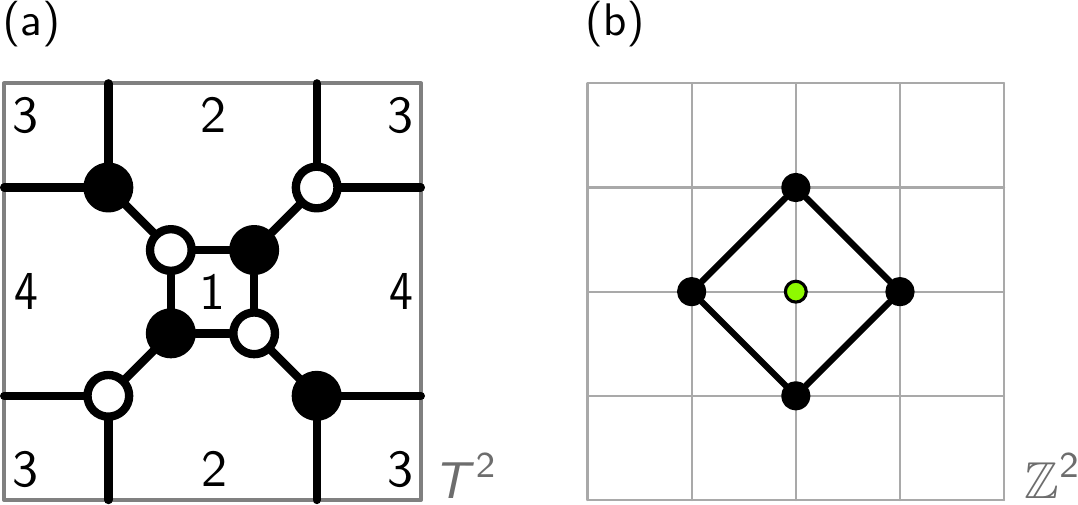} 
}
\caption{
(a) The brane tiling for the second phase of the zeroth Hirzebruch surface $F_0$, and (b) its corresponding toric diagram \cite{hirzebruch1968singularities,brieskorn1966beispiele,Morrison:1998cs,Feng:2000mi}.
\label{f_fig01}}
 \end{center}
 \end{figure}

This class of $4d$ $\mathcal{N}=1$ supersymmetric gauge theories can be represented in terms of a T-dual Type IIB brane configuration known as a brane tiling \cite{Franco:2005rj,Hanany:2005ve,Franco:2005sm}.
\tref{t_brane} summarizes the Type IIB brane configuration.
Brane tilings can be illustrated in terms of bipartite graphs on a 2-torus $T^2$ \cite{2003math.....10326K,kasteleyn1967graph} and encapsulate both the field theory information and the information about the associated toric Calabi-Yau geometry. 
\fref{f_fig01} shows an example of a brane tiling and its associated toric Calabi-Yau 3-fold, which is in this case the cone over the zeroth Hirzebruch surface $F_0$ \cite{hirzebruch1968singularities,brieskorn1966beispiele,Morrison:1998cs,Feng:2000mi}.
The mesonic moduli spaces \cite{Witten:1993yc,Benvenuti:2006qr,Feng:2007ur,Butti:2007jv} formed by the mesonic gauge invariant operators of these $4d$ $\mathcal{N}=1$ supersymmetric gauge theories with $U(1)$ gauge groups is precisely the associated toric Calabi-Yau 3-folds.
When all the gauge groups of the $4d$ $\mathcal{N}=1$ supersymmetric gauge theory are $U(N)$, then the mesonic moduli space is given by the $N$-th symmetric product of the toric Calabi-Yau 3-fold.

The gravity dual of the $4d$ worldvolume theories is Type IIB string theory on $AdS_5\times Y_5$, where $Y_5$ is the Sasaki-Einstein 5-manifold that forms the base of the associated toric Calabi-Yau 3-fold \cite{Maldacena:1997re,Morrison:1998cs,Acharya:1998db,Martelli:2004wu,Benvenuti:2004dy,Benvenuti:2005ja,Butti:2005sw}.
These $4d$ $\mathcal{N}=1$ supersymmetric gauge theories are known to flow at low energies to a superconformal fixed point. 
Under a procedure known as $a$-maximization \cite{Intriligator:2003jj,Butti:2005vn,Butti:2005ps}, the superconformal $R$-charges of the $4d$ theory are determined.
This procedure, involves the maximization of the trial $a$-charge, which takes the form
\beal{es2a50}
a(R;Y_5)
= 
\frac{3}{32} (3 \text{Tr} R^3 - \text{Tr} R) ~.~
\eea
The maximization procedure gives the value of the central charge of the superconformal field theory at the conformal fixed point.

Under the AdS/CFT correspondence \cite{Maldacena:1997re,Morrison:1998cs,Acharya:1998db}, the central charge is directly related to the minimized volume of the corresponding Sasaki-Einstein 5-manifold $Y_5$ \cite{Gubser:1998vd,Henningson:1998gx}. We have,
\beal{es2a55}
a(R;Y_5) = \frac{\pi^3 N^2}{4V(R;Y_5)}
~,~
\eea
where the R-charges $R$ and as a result the volume function $V(R;Y_5)$ can be expressed in terms of Reeb vector components $b_i$ of the corresponding Sasaki-Einstein 5-manifold \cite{Martelli:2006yb,Martelli:2005tp}.
We can reverse the statement saying that computing the minimum volume,
\beal{es2a56}
V_{min} = \text{min}_{b_i } ~V(b_i; Y_5) ~,~
\eea
is equivalent to obtaining the maximum value of the central charge $a(R; Y_5)$.
This correspondence is true for all $4d$ theories living on a stack of $N$ D3-branes probing toric Calabi-Yau 3-folds and has been checked extensively in various examples \cite{Intriligator:2003jj,Butti:2005vn,Butti:2005ps}. 

In this work, we will focus on the toric Calabi-Yau 3-folds and the corresponding Sasaki-Einstein 5-manifold $Y_5$, with particular emphasis on the minimum volume $V_{min}$ of the Sasaki-Einstein 5-manifolds $Y_5$.
Building on the pioneering work of \cite{Krefl:2017yox},
this work proposes the use of more advanced machine learning techniques.
In particular, we introduce machine learning regularization by using the
Least Absolute Shrinkage and Selection Operator (Lasso) \cite{tibshirani1996regression} in order to identify an explicit formula for the minimum volume $V_{min}$ for Sasaki-Einstein 5-manifolds $Y_5$.
We expect to be able to write the minimum volume formula in terms of features obtained from the toric diagram of the corresponding toric Calabi-Yau 3-folds.
The use of machine learning regularization allows us to eliminate parameters, reducing the necessary parameters for the volume formula to a manageable amount that is interpretable,  presentable and reusable.

Before discussing these machine learning techniques, let us first review in the following section the computation of the volume functions for toric Calabi-Yau 3-folds using Hilbert series.
\\

\section{Hilbert Series and Calabi-Yau Volumes}\label{sback3}

Given $\mathcal{X}$ as a cone over a projective variety $X$, where $X$ is realized as an affine variety in $\mathbb{C}$, the Hilbert series \cite{Benvenuti:2006qr,Feng:2007ur} is the generating function for the dimension of the graded pieces of the coordinate ring
\beal{es3a10}
\mathbb{C}[x_1,\dots, x_k] / \langle f_i \rangle ~,~
\eea
where $f_i$ are the defining polynomials of $X$. 
Accordingly, the Hilbert series takes the general form 
\beal{es3a11}
g(t;\mathcal{X}) = \sum_{i=0}^{\infty} \text{dim}_{\mathbb{C}}(X_i) t^i ~.~
\eea

For $4d$ $\mathcal{N}=1$ supersymmetric gauge theories given by brane tilings \cite{Franco:2005rj,Hanany:2005ve,Franco:2005sm}, 
we have an associated toric Calabi-Yau 3-fold $\mathcal{X}$, 
which becomes the mesonic moduli space \cite{Witten:1993yc,Benvenuti:2006qr,Feng:2007ur,Butti:2007jv} of the $4d$ $\mathcal{N}=1$ supersymmetric gauge theory when the gauge groups are all $U(1)$. 
The corresponding Hilbert series is the generating function of mesonic gauge invariant operators that form the mesonic moduli space.
For the purpose of the remaining discussion, we will consider the $4d$ $\mathcal{N}=1$ supersymmetric gauge theories given by brane tilings as abelian theories with $U(1)$ gauge groups.

Following the forward algorithm for brane tilings \cite{Feng:2000mi}, we can use GLSM fields \cite{Witten:1993yc} given by perfect matchings $p_\alpha$ \cite{Hanany:2005ve, Franco:2005rj} of the brane tilings in order to express the mesonic moduli space of the abelian $4d$ $\mathcal{N}=1$ supersymmetric gauge theory as the following symplectic quotient,
\beal{es3a20}
\mathcal{X} = {}^{\text{Irr}}\mathcal{F}^{\flat} // Q_D = \left( \mathbb{C}[p_\alpha]// Q_F \right) // Q_D ~,~
\eea
where ${\text{Irr}}\mathcal{F}^{\flat}$ is the largest irreducible component, also known as the coherent component, of the master space $\mathcal{F}^\flat$ \cite{Hanany:2010zz,Forcella:2008bb,Forcella:2008eh} of the $4d$ $\mathcal{N}=1$ supersymmetric gauge theory.
The master space is the spectrum of the coordinate ring generated by the chiral fields encoded in $p_\alpha$ and quotiented by the F-term relations encoded in $Q_F$.
In \eref{es3a20}, $Q_F$ is the $F$-term charge matrix summarizing the $U(1)$ charges originating from the $F$-terms, and $Q_D$ is the $D$-term charge matrix which summarizes the $U(1)$ gauge charges on perfect matchings $p_\alpha$. 

Following the symplectic quotient description of the mesonic moduli space in \eref{es3a20}, 
the Hilbert series can be obtained by solving the Molien integral \cite{Pouliot:1998yv},
\beal{es3a21}
g(y_\alpha; \mathcal{X}) &=& \prod_{i=1}^{c-2} \oint_{|z_i|=1} 
\frac{\ud z_i}{2\pi i z_i} 
\nn\\
&&
\times
\prod_{\alpha=1}^{c} 
\frac{1}{1-y_\alpha \prod_{j=1}^{c-3} z_j^{(Q_t)_{j\alpha}}}~,~
\eea
where $c$ is the number of perfect matchings in the brane tiling
and $Q_t = (Q_F, Q_D)$ is the total charge matrix.

\cite{Martelli:2006yb,Martelli:2005tp} showed that the same Hilbert series can be obtained directly from 
the toric diagram $\Delta$ of the toric Calabi-Yau 3-fold $\mathcal{X}$.
Given that the toric diagram $\Delta$ is a convex lattice polygon on $\mathbb{Z}^2$ with an ideal triangulation $\mathcal{T}(\Delta)$ into unit sub-triangles $\Delta_i \in \mathcal{T}(\Delta)$, the Hilbert series of the corresponding toric Calabi-Yau 3-fold $\mathcal{X}$ can be written as 
\beal{es3a30}
g(t_i; \mathcal{X}) = 
\sum_{i=1}^{r} 
\prod_{j=1}^{n}
\frac{1}{(1-\mathbf{t}^{\mathbf{u}_{i,j}})}
~,~
\eea
where $i=1,\dots,r$ is the index for the $r$ unit triangles $\Delta_i \in \mathcal{T}(\Delta)$, and $j=1,2,3$ is the index for the $3$ boundary edges of each unit triangle $\Delta_i$. 
For each boundary edge $e_j \in \Delta_i$, we have a $3$-dimensional outer normal vector $\mathbf{u}_{i,j}$ whose components are assigned the following product of fugacities, 
\beal{es3a31}
\mathbf{t}^{\mathbf{u}_{i,j}} = \prod_{a}^{3} t_a^{\mathbf{u}_{i,j}(a)} ~,~
\eea
where $\mathbf{u}_{i,j}(a)$ indicates the $a$-th component of $\mathbf{u}_{i,j}$.
We note that $\mathbf{u}_{i,j}$ is a $3$-dimensional vector because the defining vertices of $\Delta$ and $\Delta_i$ are all on a plane at height $z=1$ such that their coordinates are of the form $(x, y, 1)$.
As a result, the vectors $\mathbf{u}_{i,j}$ corresponding to edge $e_j \in \Delta_j$ are normal to the 3-dimensional surface given by the vectors connecting the origin $(0,0,0)$ to the two bounding vertices of $e_j \in \Delta_j$.

\begin{figure}[ht!!]
\begin{center}
\resizebox{0.95\hsize}{!}{
\includegraphics[height=5cm]{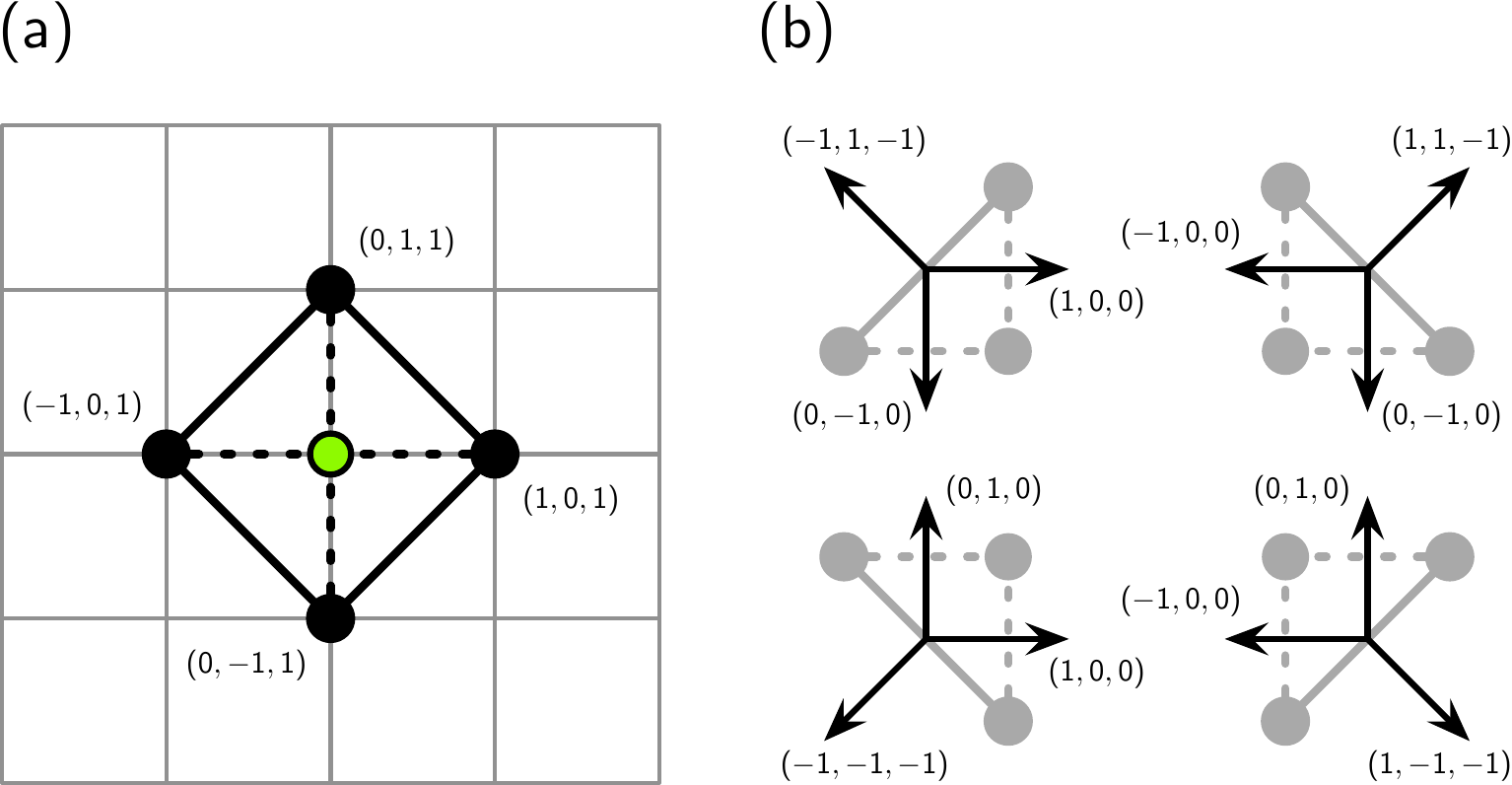} 
}
\caption{
(a) The triangulated toric diagram for the zeroth Hirzebruch surface $F_0$, and (b) the corresponding normal vectors $\mathbf{u}_{i,j}$ for each unit triangle $\Delta_i$ in the triangulation. 
\label{f_fig02}}
 \end{center}
 \end{figure}

It is important to note that the fugacities $t_1, t_2, t_3$ in \eref{es3a31} relate to the components of normal vectors $\mathbf{u}_{i,j}$, and therefore depend on the triangulation and the particular instance in a given $GL(2,\mathbb{Z})$ toric orbit of a toric diagram on the $z=1$ plane. 
In comparison, the fugacities $y_\alpha$ in \eref{es3a21} refer to the GLSM fields $p_\alpha$ given by perfect matchings of the corresponding brane tiling. 
Since perfect matchings can be mapped directly to chiral fields in the $4d$ $\mathcal{N}=1$ supersymmetric gauge theory, the fugacities $y_\alpha$ in \eref{es3a21} can be mapped to fugacities counting global symmetry charges carried by chiral fields in the $4d$ theory.
Because both Hilbert series from \eref{es3a21} and \eref{es3a30} refer to the same toric Calabi-Yau 3-fold $\mathcal{X}$, there exists a fugacity map between $y_\alpha$ and $t_1, t_2, t_3$ that identifies the two Hilbert series with each other. 

For the rest of the discussion, let us consider Hilbert series for toric Calabi-Yau 3-folds $\mathcal{X}$ that are in terms of fugacities $t_1, t_2, t_3$ corresponding to coordinates of the normal vectors $\mathbf{u}_{i,j} \in \mathbb{Z}^3$ of the toric diagram $\Delta$.
Given the Hilbert series $g(t_i; \mathcal{X})$, we can obtain the volume function \cite{Martelli:2006yb,Martelli:2005tp} of the Sasaki-Einstein 5-manifold $Y_5$ using,
\beal{es3a40}
V(b_i; Y_5) = \lim_{\mu \rightarrow 0} \mu^3 g(t_i = \exp[-\mu b_i]; \mathcal{X}) ~,~
\eea
where $b_i$ are the Reeb vector components with $i=1, \dots 3$. 
We note that the Reeb vector ${\bf b}=(b_1,b_2,b_3)$ is always in the interior of the toric diagram $\Delta$ and can be chosen such that one of its components is set to
\beal{es3a41}
b_3  = 3~,~
\eea
for toric Calabi-Yau 3-folds $\mathcal{X}$.
We further note that the limit in \eref{es3a40} takes the leading order in $\mu$ in the expansion for $g(t_i = \exp[-\mu b_i]; \mathcal{X})$, which is shown to refer to the volume of the Sasaki-Einstein base $Y_5$ in \cite{Martelli:2006yb,Martelli:2005tp}.

Let us consider in the following paragraph an example of the computation of the volume function in terms of Reeb vector components $b_i$ for the Sasaki-Einstein base of the cone over the zeroth Hirzebruch surface $F_0$ \cite{hirzebruch1968singularities,brieskorn1966beispiele,Morrison:1998cs,Feng:2000mi}.
\\

\noindent
\textit{Example: $F_0$.}
The toric diagram, its triangulation and the outer normal vectors $\mathbf{u}_{i,j}$ for the cone over the zeroth Hirzebruch surface $F_0$ \cite{hirzebruch1968singularities,brieskorn1966beispiele,Morrison:1998cs,Feng:2000mi} are shown in \fref{f_fig02}(a).
The cone over the zeroth Hirzebruch surface $F_0$ is an interesting toric Calabi-Yau 3-fold because it has two distinct corresponding $4d$ $\mathcal{N}=1$ supersymmetric gauge theories represented by two distinct brane tilings that are related by Seiberg duality \cite{Seiberg:1994pq,2001JHEP...12..001B,Feng:2000mi}. One of the brane tilings is shown in \fref{f_fig01}.

Using the outer normal vectors $\mathbf{u}_{i,j}$ for each of the four unit sub-triangles $\Delta_i$ of the toric diagram for $F_0$ in \fref{f_fig02}(b), 
we can use \eref{es3a30} to write down the Hilbert series,
\beal{es3a50}
&&
g(t_i; F_0) 
= 
\frac{1}{(1-t_1)(1-t_2^{-1})(1-t_1^{-1} t_2 t_3^{-1})}
\nn\\
&& \hspace{0.5cm}
+
\frac{1}{(1-t_1^{-1})(1-t_2^{-1})(1-t_1 t_2 t_3^{-1})}
\nn\\
&& \hspace{0.5cm}
+
\frac{1}{(1-t_1)(1-t_2)(1-t_1^{-1} t_2^{-1} t_3^{-1})}
\nn\\
&& \hspace{0.5cm}
+
\frac{1}{(1-t_1^{-1})(1-t_2)(1-t_1 t_2^{-1} t_3^{-1})}
~.~
\eea

Using the limit in \eref{es3a40}, we can derive the volume function of the Sasaki-Einstein base directly from the Hilbert series as follows,
\beal{es3a51}
&&
V(b_i; F_0)=
\nn\\
&&
\frac{24}{
(b_1 - b_2 - 3) (b_1 - b_2 + 3) (b_1 + b_2 - 3) (b_1 + b_2 + 3) 
}
~,~
\nn\\
\eea 
where $b_3 = 3$.
When we find the global minimum of the volume function $V(b_i; F_0)$, we obtain
\beal{es3a52}
V_{min} =  \text{min}_{b_i } ~V(b_i; F_0)  = \frac{8}{27} \simeq 0.29630
~,~
\eea
up to 5 decimal points, which occurs at critical Reeb vector components $b_1^* =  b_2^* = 0$.
In the remainder of this work, we will maintain a precision level of 5 decimal points for all numerical measurements.
\\

\section{Features of Toric Diagrams and Regression}\label{sback3}

The aim of this work is to identify an expression for the minimum volume $V_{min}$ of Sasaki-Einstein 5-manifolds $Y_5$ 
in terms of parameters that we know from the corresponding toric Calabi-Yau 3-folds $\mathcal{X}$.
We refer to these parameters as features, denoted as $x_a$, of the toric Calabi-Yau 3-fold $\mathcal{X}$.

Assuming that we have $N_x$ features $x_a$ for a given toric Calabi-Yau 3-fold, the proposal in \cite{Krefl:2017yox} states that we can write down a candidate linear function
for the inverse minimum volume
in terms of these features as follows, 
\beal{es4a10}
1 / \hat{V}_{min} (x^j_a) \equiv \hat{y}^j = \beta_0 + \sum_{a=1}^{N_x} \beta_a x_a^j ~,~
\eea
where $\beta_0$ and $\beta_a$ are real coefficients, and $j$ labels the particular toric Calabi-Yau 3-fold $\mathcal{X}^j$ with its corresponding toric diagram $\Delta^j \in \mathbb{Z}^2$.

Let us refer to the inverse of the actual minimum volume obtained by volume minimization as $1/V_{min}^j \equiv y^j$ for a given toric Calabi-Yau 3-fold $\mathcal{X}^j$.
If for a set $S$ of $N=|S|$ toric Calabi-Yau 3-folds $\mathcal{X}^j$, we know the actual minimum volumes $V_{min}^j$ via volume minimization, then we can calculate the following residual sum of squares of the difference between the inverses of the actual and the expected minimum volumes for the entire set $S$, 
\beal{es4a15}
\mathcal{L} 
&=&
\frac{1}{2N} 
\sum_{j=1}^{N=|S|} 
\left(
y^j - \hat{y}^j
\right)^2 
\nn\\
&=& 
\frac{1}{2N} 
\sum_{j=1}^{N}
\left(
1/V_{min}^j - \beta_0 - \sum_{a=1}^{N_x} \beta_a x_a^j
\right)^2 
~.~
\nn\\
\eea
Here, $\mathcal{L}$ can be considered as a loss function \cite{goodfellow2016deep} that evaluates the performance of the candidate function for the minimum volume in \eref{es4a10}.
In multiple linear regression \cite{gauss1823theoria, fisher1922mathematical,mendenhall2003second,  freedman2009statistical,jobson2012applied}, as initially proposed in \cite{Krefl:2017yox}, the optimization task is to minimize the loss function in \eref{es4a15} for a given dataset $S$ of toric Calabi-Yau 3-folds, 
\beal{es4a16}
\text{argmin}_{\beta_0, \beta_a} \mathcal{L} ~.~
\eea

In \cite{Krefl:2017yox}, multiple linear regression was used to obtain a candidate minimum volume function using the following feature set, 
\beal{es4a20}
x_a^j \in \{ f_1, f_2, f_3, f_1 f_2, f_1 f_3 , \dots, f_1^2, f_2^2, f_3^2 \}^j~,~
\eea
where 
\beal{es4a21}
f_1 = I ~,~
f_2 = E ~,~
f_3 = V ~,~
\eea
corresponding respectively to the number of internal lattice points in $\Delta^j$,
the number of boundary lattice points in $\Delta^j$, and
the number of vertices that form the extremal corner points in $\Delta^j$,
for a given toric Calabi-Yau 3-fold $\mathcal{X}^j$.
Under \textit{Pick's theorem} \cite{pick1899geometrisches},
these features are related as follows,
\beal{es4a22}
A= I + E/2 - 1 ~,~
\eea
where $A$ is the area of the toric diagram $\Delta$, with the area of the smallest unit triangle in $\mathbb{Z}^2$ having $A=1/2$.

With a dataset $S$ of $N=15,147$ toric Calabi-Yau 3-folds, the work in \cite{Krefl:2017yox} showed that the candidate linear function in \eref{es4a10} with features given by \eref{es4a20} is able to estimate the inverse minimum volume with an expected percentage relative error of 2.2\%.
In this work, we expand upon the accomplishments of \cite{Krefl:2017yox} by introducing novel features that describe toric Calabi-Yau 3-folds, augmenting the datasets for toric Calabi-Yau 3-folds, and applying machine learning techniques incorporating regularization.
These improvements are designed to address some of the shortcomings of the work in \cite{Krefl:2017yox} as well as give explicit interpretable formulas for the minimum volume for toric Calabi-Yau 3-folds. 
\\

\begin{figure}[ht!!]
\begin{center}
\resizebox{0.78\hsize}{!}{
\includegraphics[height=5cm]{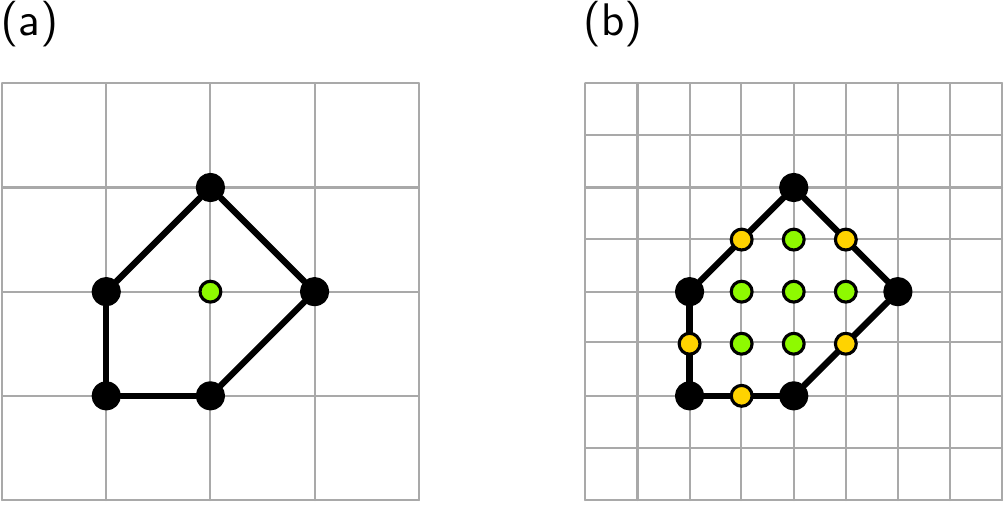} 
}
\caption{
(a) The toric diagram $\Delta_1$ for the cone over $\text{dP}_1$, and (b) the corresponding $2$-enlarged toric diagram $\Delta_2$ with $n=2$.
\label{f_fig03}}
 \end{center}
 \end{figure}

\noindent
\textit{New Features.}
We introduce several new features that describe a toric Calabi-Yau 3-fold and are obtained from the corresponding toric diagram $\Delta$.
By defining the $n$-enlarged toric diagram as,
\beal{es4a30}
\Delta_n = \{ nv = (nx, ny) ~|~ v=(x,y) \in \Delta \} ~,~
\eea
where $n\in \mathbb{Z}^+$ and $v=(x,y)\in \mathbb{Z}^2$ are the coordinates of the vertices in the original toric diagram $\Delta$.
We note that $\Delta_1 = \Delta$.
These $n$-enlarged toric diagrams $\Delta_n$ also appeared in \cite{Berglund:2021ztg} for the study of Hodge numbers of Calabi-Yau manifolds that are constructed as hypersurfaces in toric varieties given by $\Delta$.

Using the $n$-enlarged toric diagram $\Delta_n^j$ for a given toric Calabi-Yau 3-fold $\mathcal{X}^j$, we can now refer to the area of $\Delta_n$ as $A_n$, the number of internal lattice points of $\Delta_n$ as $I_n$, and the number of boundary lattice points in $\Delta_n$ as $E_n$.
We further note that the number of vertices $V_n$ corresponding to extremal corner points in $\Delta_n$ is the same for $V$ in $\Delta$ for all $n$, i.e. $V_n = V$.

In our work, we use features of a toric Calabi-Yau 3-fold $\mathcal{X}^j$ that are composed from members of the following set,
\beal{es4a30b}
\{ A,V, E, I_n\}^j ~,~
\eea
where $n=1,\dots,7$.
These are defined through the corresponding toric diagram $\Delta^j$ and its corresponding $n$-enlarged toric diagram $\Delta_n^j$.
Through the application of machine learning regularization, our objective is to differentiate between features that contribute to the expression for the minimum volume associated with a toric Calabi-Yau 3-fold and those that do not.
\\

\begin{table}[t]
\centering
\begin{tabular}{|c|c|c|}
\hline
Set & Description & $|S_m|$
\\
\hline \hline
$S_\text{1a}$ & all polytopes $5 \times 5$ lattice box  & 15,327 \\
$S_\text{1b}$ & all polytopes $r=3.5$ circle & 31,324 \\
\hline
$S_\text{2a}$ &selected polytopes $30 \times 30$ lattice box & 202,015\\
$S_\text{2b}$ & selected polytopes $r=15$ circle & 201,895\\
\hline
\end{tabular}
\caption{For training the machine learning models, we make use of 4 sets $S_m$ of toric diagrams with different sizes $|S_m|$.}
\label{t_data01}
\end{table}

\begin{figure}[ht!!]
\begin{center}
\resizebox{\hsize}{!}{
\includegraphics[height=5cm]{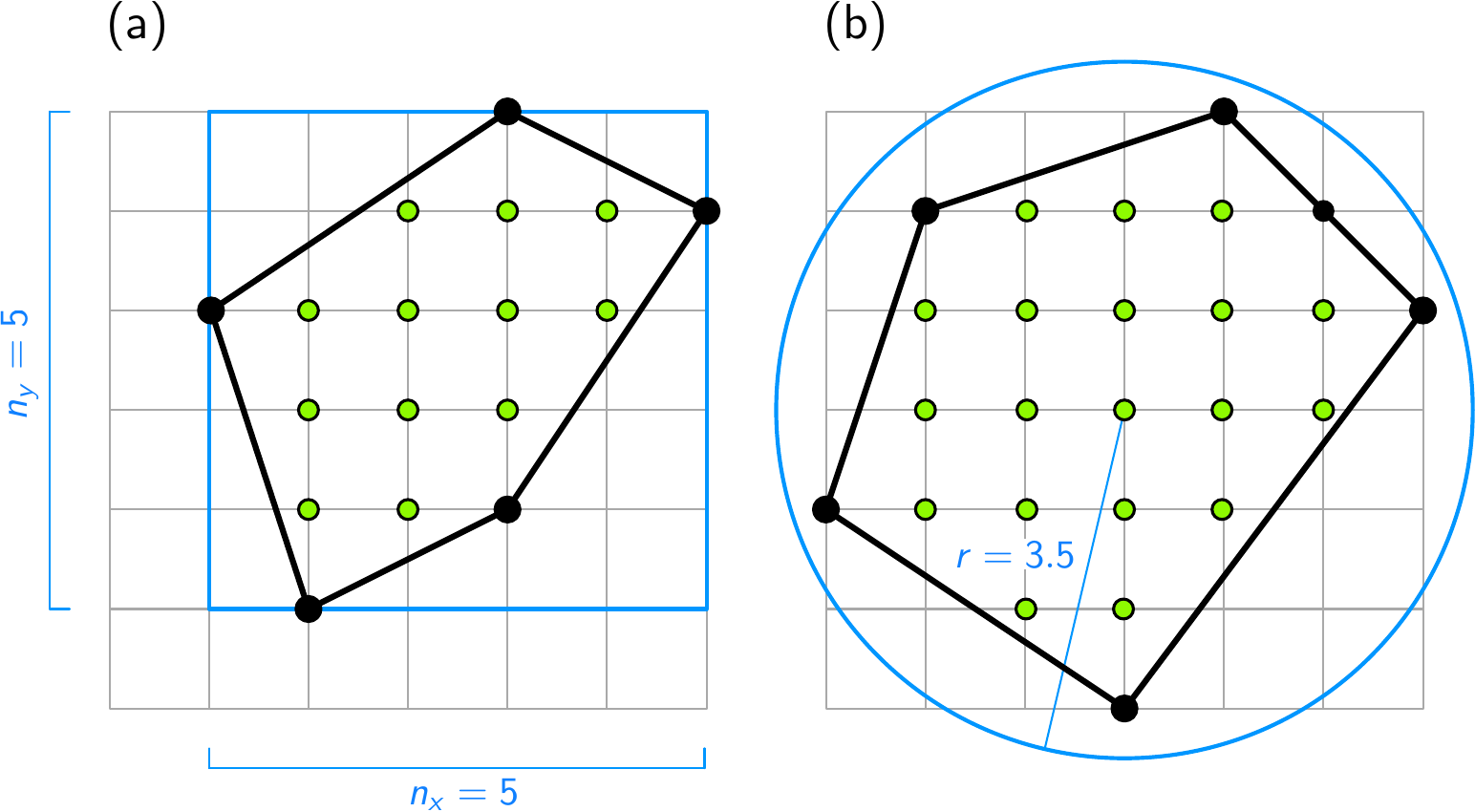} 
}
\caption{
(a) Toric diagrams in datasets $S_\text{1a}$ and $S_\text{2a}$ are constrained by a $n_x\times n_y$ lattice box in $\mathbb{Z}^2$, whereas (b) toric diagrams in datasets $S_\text{1b}$ and $S_\text{2b}$ are constrained by a circle of radius $r$ with the center at $(0,0)\in\mathbb{Z}^2$. 
\label{f_fig04}}
 \end{center}
 \end{figure}

\begin{figure*}[ht!!]
\begin{center}
\resizebox{\hsize}{!}{
\includegraphics[height=5cm]{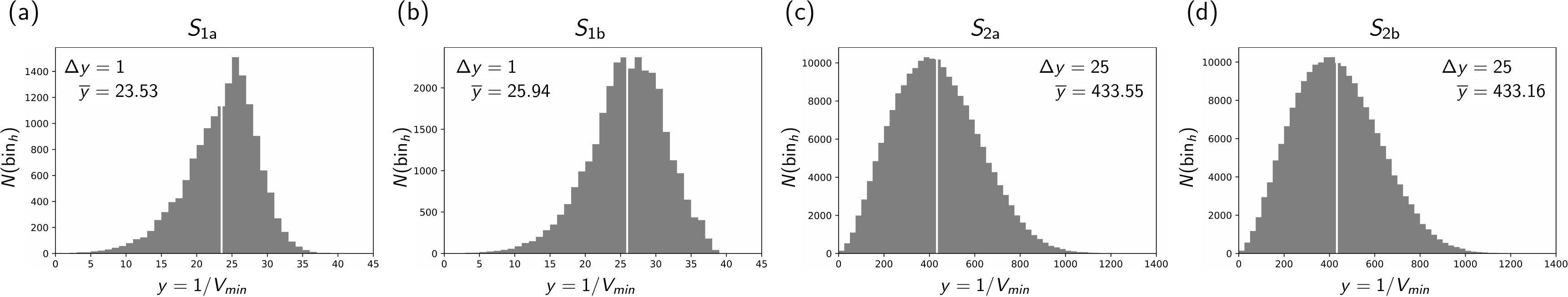} 
}
\caption{
The distribution of expected minimum volumes $y=1/V_{min}$ for the datasets (a) $S_\text{1a}$, (b) $S_\text{1b}$, (c) $S_\text{2a}$ and (c) $S_\text{2b}$.
The mean expected value $\overline{y}$ is indicated by a white line.
The histograms for values of $y=1/V_{min}$ are obtained for bin sizes $\Delta y$ with the number of toric diagrams in $\text{bin}_h$ given by $N(\text{bin}_h)$.
\label{f_fig05}}
 \end{center}
 \end{figure*}

\noindent
\textit{New Sets of Toric Calabi-Yau 3-folds.}
The aim of this work is to make use of machine learning with regularization in order to identify an interpretable formula that accurately estimates the minimum volume of Sasaki-Einstein 5-manifolds corresponding to toric Calabi-Yau 3-folds.
The interpretability of the minimum volume formula is achieved by the lowest possible number of features on which the formula depends on.
In order to train such a regularized machine learning model, we establish four sets $S_m$ of toric Calabi-Yau 3-folds $\mathcal{X}^j$, for which the corresponding minimum volumes are known.
These sets $S_m$ are defined as follows:
\begin{itemize}
\item \textit{$S_\text{1a}$:} This set consists of toric Calabi-Yau 3-folds whose toric diagrams fit into a $5\times 5$ lattice box in $\mathbb{Z}^2$ as illustrated in \fref{f_fig04}(a). 
This set contains a certain degree of redundancy given that convex lattice polygons related by a $GL(2,\mathbb{Z})$ transformation on their vertices refer to the same toric Calabi-Yau 3-fold.
Accordingly, we restrict ourselves to toric diagrams $\Delta^j$ that give unique combinations of the form $(1/V_{min}^j, V^j, E^j, I^j)$.
This results in a dataset of $|S_\text{1a}|=15,327$ distinct toric diagrams with unique inverse minimum volumes $1/V_{min}^j$ up to 6
decimal points.
\item \textit{$S_\text{1b}$:} The second set consists of toric Calabi-Yau 3-folds whose toric diagrams fit inside a circle centered at the origin $(0,0)$ on the $\mathbb{Z}^2$ lattice with radius $r=3.5$ as illustrated in \fref{f_fig04}(b).
By imposing the condition that we want $GL(2,\mathbb{Z})$-distinct toric diagrams $\Delta^j$ with unique combinations of the form $(1/V_{min}^j, V^j, E^j, I^j)$, we obtain $|S_\text{1b}|=31,324$ toric diagram for this set.

\item \textit{$S_\text{2a}$:} For this set, we choose randomly 300,000 toric diagrams that fit into a $30\times 30$ lattice box in $\mathbb{Z}^2$.
By imposing the condition that the toric diagrams $\Delta^j$ have unique combinations of the form $(1/V_{min}^j, V^j, E^j, I^j)$, we obtain $|S_\text{2a}|=202,015$ toric diagram for this set.

\item \textit{$S_\text{2b}$:} For this set, we choose randomly 300,000 toric diagrams that fit into a circle centered at the origin $(0,0)$ on the $\mathbb{Z}^2$ lattice with radius $r=15$. By imposing the condition that the toric diagrams $\Delta^j$ have unique combinations of the form $(1/V_{min}^j, V^j, E^j, I^j)$, we obtain $|S_\text{2b}|=201,895$ toric diagram for this set.
\end{itemize}
The distribution of inverse minimum volumes $1/V_{min}$ for the above sets of toric diagrams is illustrated together with the mean inverse minimum volume $\overline{y} = \langle 1/V_{min} \rangle = \frac{1}{|S_m|} \sum_{j=1}^{|S_m|} 1/V_{min}^j$ in \fref{f_fig05}. 
In the following sections, we make use of regularized machine learning in order to identify functions that optimally estimate the inverse minimum volume $1/V_{min}$ in each of the above datasets. 
\\

\noindent
\textit{Machine Learning Models and Regularization.}
In order to obtain a function for the minimum volume of Sasaki-Einstein 5-manifolds corresponding to toric Calabi-Yau 3-folds in terms of features obtained from the corresponding toric diagrams, we make use of the following machine learning models: 
\begin{itemize}
\item \textbf{Polynomial Regression (PR).}
We make use of polynomial regression \cite{montgomery2021introduction}, where the relationship between the feature variables $x_a^j$ and the predicted variable $\hat{y}^j$, is given by
\beal{es3a50}
\hat{y}^j = 
\beta_0 + \sum_{a=1}^{N_x} \beta_{a} x_a^j  ~.~
\eea
Here, $\beta_0$ and $\beta_{a}$ are real coefficients, $N_x$ is the number of features, and $j$ labels the particular sample in the data set that is used to train this machine learning model.
In our case, the data set consists of toric Calabi-Yau 3-folds $\mathcal{X}^j$, where the corresponding minimum volume $V_{min}^j$ is given by $y=1/V_{min}^j$.
Here we note that the features $x_a^j$ are taken from the set $\{ (f_u^j)^a (f_v^j)^b ~|~  1 \leq a+b \leq 2,~ a,b\in\mathbb{Z}^+ \}$ with $f_u^j \in \{A, V, E,I_n\}^j$, where $n=1,\dots,7$.

\item \textbf{Logarithmic Regression (LR).}
We make use of logarithmic regression \cite{montgomery2021introduction} in order to help linearize relationships between features $x_a^j$
that are potentially multiplicative in their contribution towards the predicted variable $\hat{y}^j$.
To be more precise, we make use of a $\log$-$\log$ model where we $\log$-transform both the predicted variable $\hat{y}^j$ and the features $x_a^j$.
The predicted variable is then given by,
\beal{es3a55}
\log (\hat{y}^j) = 
\beta_0 + \sum_{a=1}^{N_x} \beta_{a} \log(x_a^j)
\eea
where $\beta_0$ and $\beta_{a}$ are real coefficients, and $N_x$ is the number of $\log$-transformed features of the form $\log(x_a^j)$. 
The label $j$ corresponds to a particular toric Calabi-Yau 3-fold $\mathcal{X}^j$ whose corresponding minimum volume $V_{min}^j$ is given by $y^j=1/V_{min}^j$.
Here we note that the $\log$-transformed features of the form $\log(x_a^j)$ are taken from the set $\{ (\log(f_u^j))^a (\log(f_v^j))^b ~|~  1 \leq a+b \leq 2,~ a,b\in\mathbb{Z}^+  \}$ with $f_u^j \in \{A, V, E, I_n\}^j$, where $n=3,\dots,7$.
Here, we do not make use of $I_1$ and $I_2$.

\end{itemize}
When we introduce regularization \cite{tikhonov1963regularization,hastie2009elements} into polynomial regression and logarithmic regression, we minimize the following loss function between the predicted variable $\hat{y}^j$ and the expected variable $y$, 
\beal{es3a60}
\mathcal{L} = \frac{1}{2N}\sum_{j=1}^{N} (y^j - \hat{y}^j)^2 ~+ \Delta\mathcal{L}
~,~
\eea
where $\Delta\mathcal{L}$ is the regularization term in the loss function.
The loss function in \eref{es3a60} is iteratively minimized during the optimization process and we set for all following computations the maximum number of iterative steps to be $N_{max} = 10,000$.
The precise form of the regularization term in the loss function as well as the different regularization schemes in machine learning are discussed in the following section.
\\

\section{Least Absolute Shrinkage and Selection Operator (Lasso) and Regularization}\label{sback4}

The Least Absolute Shrinkage and Selection Operator (Lasso) \cite{tibshirani1996regression} is a machine learning regularization technique primarily employed to prevent overfitting in supervised machine learning. 
However, it can also be utilized for feature selection.
In our work, 
the overarching goal in employing Lasso is to introduce a machine learning model capable of delivering optimal predictions for the minimum volume for toric Calabi-Yau 3-folds while using the fewest features from the training dataset.
For problems such as the one considered in this work, 
it is quintessential to be able to obtain formulas with a small number of parameters. 
As a result, using Lasso is particularly suited for discovering new mathematical formulas such as the one aimed for in this work for the minimum volume for toric Calabi-Yau 3-folds.

In the following section, we give a brief overview of several regularization schemes including Lasso in the context of supervised machine learning for the minimum volume formula for toric Calabi-Yau 3-folds.
\\

\noindent
\textit{Regularization.}
Regularization in machine learning is a technique usually used to avoid overfitting the dataset during model training.
This is done by adding a penalty term in the loss function.
The introduction of the added regularization term $\Delta \mathcal{L}$, resulting in an updated loss function of the form,
\beal{es5a10}
\mathcal{L} + \Delta \mathcal{L}~,~
\eea
serves the purpose of constraining the possible parameter values within the supervised machine learning model.
In the case of multiple linear regression as first introduced in \cite{Krefl:2017yox} and reviewed in section \sref{sback3},
these parameters would be the real coefficients $\beta_0$ and $\beta_a$ in the candidate linear function in \eref{es4a10} for  the expected minimum volume given by $\hat{y}^j = 1/\hat{V}^j_{min}$.
By restricting the values for these parameters, regularization effectively makes it harder for the supervised machine learning model to give a candidate function for the minimum volume $V_{min}$ with many terms in the function.
This prevents the machine learning model to overfit the dataset of minimized volumes for toric Calabi-Yau 3-folds.

Let us review the following three regularization schemes:
\begin{itemize}
\item \textbf{L1 Regularization (Lasso).}
This regularization scheme also known as Least Absolute Shrinkage and Selection Operator (Lasso) \cite{tibshirani1996regression} adds the following linear regularization term to the loss function of the regression model, 
\beal{es5a20}
\Delta \mathcal{L}_{\text{L1}} = \alpha \sum_{a=1}^{N_x} |\beta_a| ~,~
\eea
where $\beta_a$ are the real parameters of the regression model.
$\alpha$ is a real regularization parameter. 
Increasing the value of $\alpha$ has the effect of increasing the strength of the L1 regularization. 

\item \textbf{L2 Regularization (Ridge).}
Another regularization scheme is known as Ridge regularization or L2 regularization \cite{hoerl1970ridge}.
It adds the following quadratic regularization term to the loss function of the regression model, 
\beal{es5a25}
\Delta \mathcal{L}_{\text{L2}} = \alpha \sum_{a=1}^{N_x} \beta_a^2 ~,~
\eea
where $\beta_a$ are the real parameters of the regression model and $\alpha$ is again the real regularization parameter.

\item \textbf{Elastic Net (L1 and L2).}
Elastic Net \cite{zou2005regularization} is a combination of L1 (Lasso) and L2 (Ridge) regularization and adds the following regularization terms to the loss function, 
\beal{es5a30}
\Delta \mathcal{L}_{\text{L1,L2}}
= 
\alpha_1 \sum_{a=1}^{N_x} |\beta_a|
+ 
\alpha_2 \sum_{a=1}^{N_x} \beta_a^2
~,~
\eea
where $\alpha_1$ and $\alpha_2$
are relative real regularization parameters that regulate the proportion of L1 regularization and L2 regularization in this regularization scheme.
\end{itemize}
Amongst these regularization schemes in supervised machine learning, we are going to mainly focus on Lasso and L1 regularization for the remainder of this work.
While all three regularization schemes share the common goal of constraining the range of values for the model parameters $\beta_a$, it is noteworthy that only Lasso possesses the unique property of inducing sparsity among the model parameters, resulting in the complete elimination of certain parameters during the training process.

\begin{figure}[ht!!]
\begin{center}
\resizebox{0.95\hsize}{!}{
\includegraphics[height=5cm]{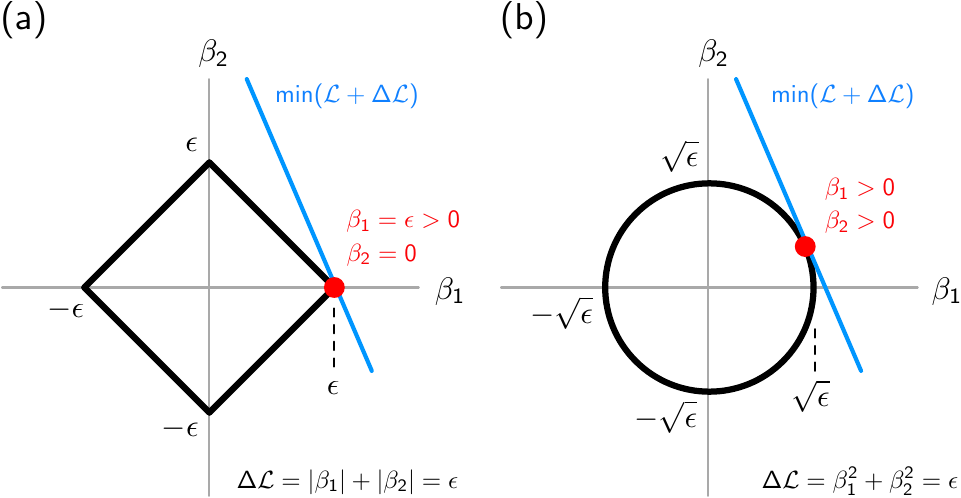} 
}
\caption{
Parametric plots for $\beta_1$ and $\beta_2$ for a 2-parameter model \cite{hastie2009elements}. (a) In L1 regularization (Lasso), the minimum of the regularized loss function $\min(\mathcal{L}+\Delta\mathcal{L})$ is more likely to be located when one of the parameters vanishes, in comparison to the case (b) in L2 regularization (Ridge) where the minimum of the regularized loss function $\min(\mathcal{L}+\Delta\mathcal{L})$ is equally more likely located at non-zero values of the parameters. 
This illustrates that L1 regularization is more suited in eliminating parameters under optimization.
\label{f_fig06}}
 \end{center}
 \end{figure}

\begin{figure*}[ht!!]
\begin{center}
\resizebox{0.9\hsize}{!}{
\includegraphics[height=5cm]{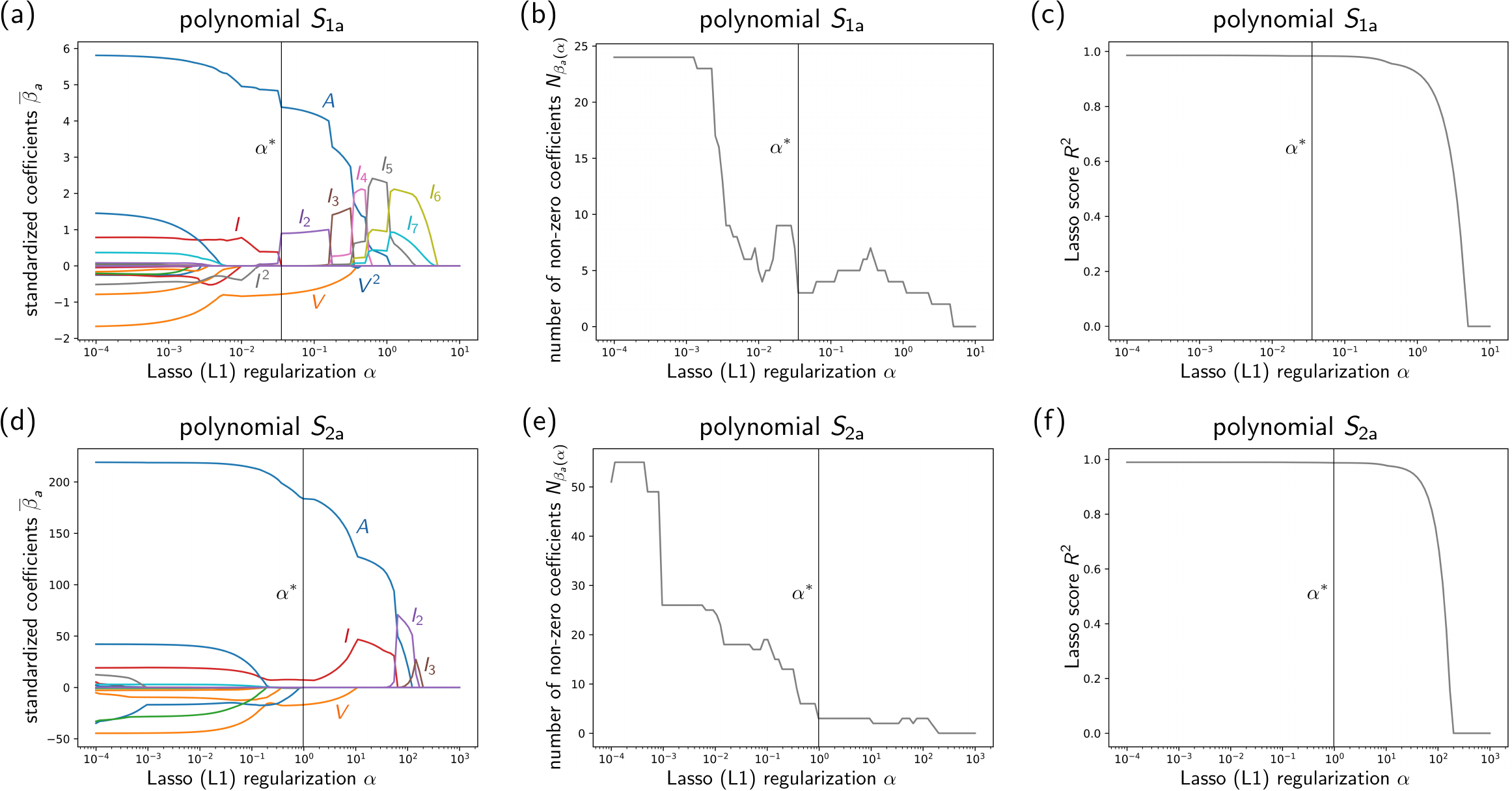} 
}
\caption{
The L1 (Lasso) regularization parameter $\alpha$ for polynomial regression on dataset $S_\text{1a}$ (15,327 toric diagrams in $5\times 5$ lattice box) against (a) the standardized coefficients $\overline{\beta}_{a}(\alpha)$, (b) the number of non-zero coefficients $N_{\beta_a(\alpha)}$, and (c) the corresponding $R^2(\alpha)$-score.
The optimal regularization parameter $\alpha^*$ was found in the range $\alpha = 10^{-4} ,\dots, 10^{1}$ by taking steps of $\Delta \alpha \simeq 1.12202$.
We also have the L1 (Lasso) regularization parameter $\alpha$ for polynomial regression on dataset $S_\text{2a}$ (202,015 random toric diagrams in $30 \times 30$ lattice box) against (c) the standardized coefficients $\overline{\beta}_a(\alpha)$, (d) the number of non-zero coefficients $N_{\beta_a(\alpha)}$, and (e) the corresponding $R^2(\alpha)$-score.
The optimal regularization parameter $\alpha^*$ was found in the range $\alpha = 10^{-4} ,\dots, 10^{3}$ by taking steps of $\Delta \alpha \simeq 1.17490$.
\label{f_fig07}}
 \end{center}
 \end{figure*}

\begin{figure*}[ht!!]
\begin{center}
\resizebox{0.9\hsize}{!}{
\includegraphics[height=5cm]{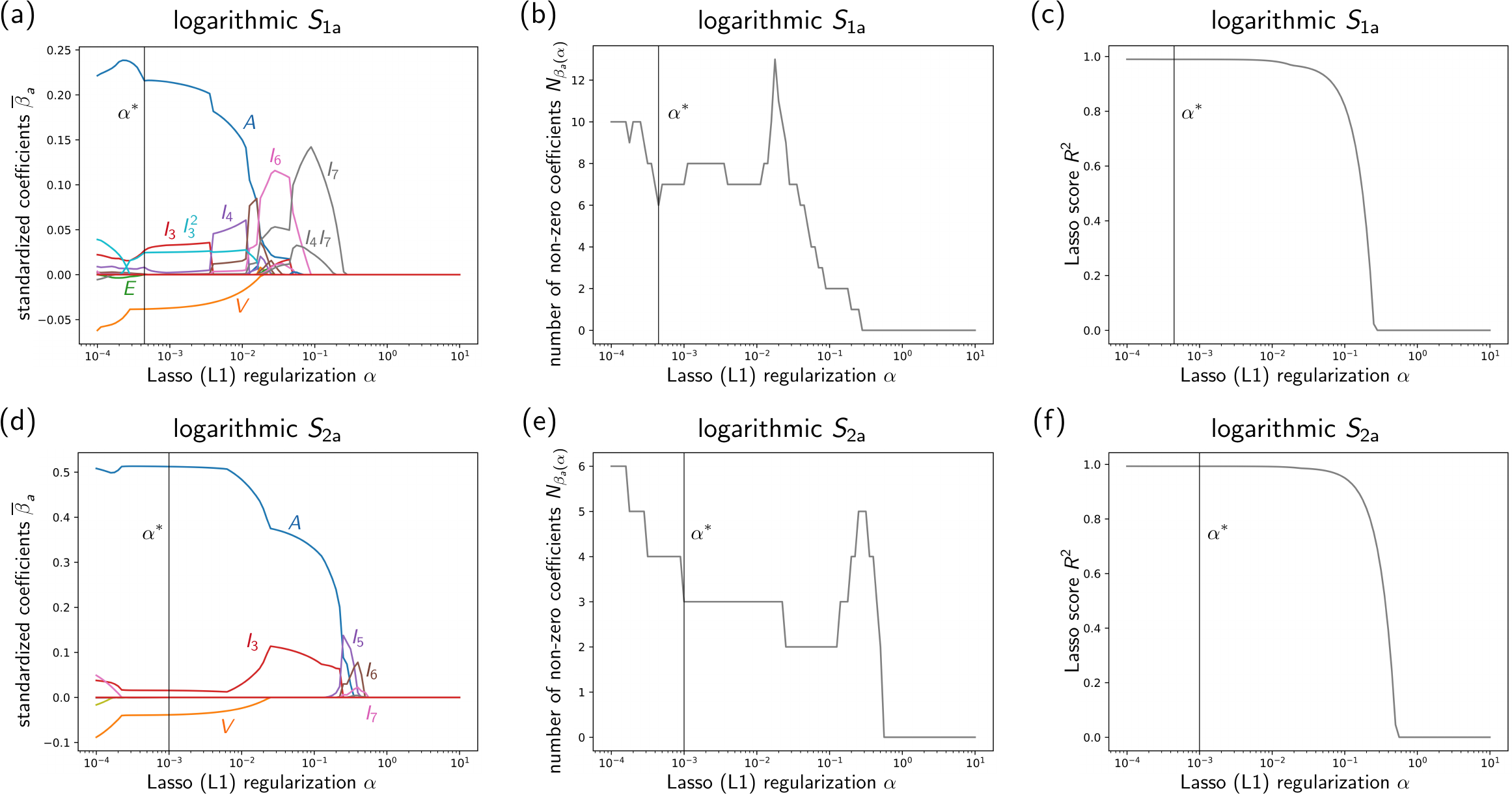} 
}
\caption{
The L1 (Lasso) regularization parameter $\alpha$ for logarithmic regression on dataset $S_\text{1a}$ (15,327 toric diagrams in $5\times 5$ lattice box) against (a) the standardized coefficients $\overline{\beta}_{a}(\alpha)$, (b) the number of non-zero coefficients $N_{\beta_a(\alpha)}$, and (c) the corresponding $R^2(\alpha)$-score.
The optimal regularization parameter $\alpha^*$ was found in the range $\alpha = 10^{-4} ,\dots, 10^{1}$ by taking steps of $\Delta \alpha \simeq 1.12202$.
We also have the L1 (Lasso) regularization parameter $\alpha$ for logarithmic regression on dataset $S_\text{2a}$ (202,015 random toric diagrams in $30 \times 30$ lattice box) against (c) the standardized coefficients $\overline{\beta}_a(\alpha)$, (d) the number of non-zero coefficients $N_{\beta_a(\alpha)}$, and (e) the corresponding $R^2(\alpha)$-score.
The optimal regularization parameter $\alpha^*$ was found in the range $\alpha = 10^{-4} ,\dots, 10^{3}$ by taking steps of $\Delta \alpha \simeq 1.17490$.
\label{f_fig08}}
 \end{center}
 \end{figure*}

There are several arguments why Lasso enables the complete elimination of some of the model parameters and the corresponding features in the candidate function for the minimum volume $V_{min}$ for toric Calabi-Yau 3-folds. 
In order to illustrate this, let us consider the case with $N_x=2$ features $x_1^j$ and $x_2^j$, for which the L1 and L2 regularization terms take respectively the following form, 
\beal{es5a35}
\Delta\mathcal{L}_{\text{L1}} = \alpha (|\beta_1| + |\beta_2|) 
~,~
\Delta\mathcal{L}_{\text{L2}} = \alpha (\beta_1^2 + \beta_2^2) 
~.~
\eea
If we assume that under optimization, the regularization terms reach a value $\Delta\mathcal{L}_{\text{L1}} = \epsilon$  and $\Delta\mathcal{L}_{\text{L2}} = \epsilon$ for $\alpha> 0$ and $\epsilon \in \mathbb{R}$, 
we can draw the parametric plots for the two regularization terms as shown in \fref{f_fig06} \cite{hastie2009elements}.
We can see from the plots in \fref{f_fig06} that for L1 regularization, the minimum of the total loss function is more likely achieved when one of the two parameters $\beta_1$ or $\beta_2$ approaches 0. 
This is in part due to the absolute values taken for the parameters in the linear L1 regularization term.

As a result, Lasso regularization is particularly suited for feature selection and parameter elimination in regression models. 
In our work, we employ L1 regularization to derive a formula for the minimum volume $V_{min}$ of Sasaki-Einstein 5-manifolds corresponding to toric Calabi-Yau 3-folds that is interpretable, presentable and reusable.
\\

\section{Candidates for Minimum Volume Functions}\label{sback5}

In this work, our aim is to apply Lasso regularization in order to identify explicit formulas for the minimum volume for toric Calabi-Yau 3-folds.
By doing so, our aim is to maximize the accuracy of the formulas that we find while minimizing the number of parameters the formulas depend on, making them interpretable and readily presentable.
\\

\begin{table*}[ht!!]
\centering
\resizebox{0.75\hsize}{!}{
\begin{tabular}{|c|c|c|c|c|}
\hline
data set & $y=1/V_{min}$ & $\alpha^*$ & $N_{\beta_a(\alpha^*)}$ & $R^2(\alpha^*)$ 
\\
\hline \hline
$S_\text{1a}$ & $\hat{y}_\text{1a}^\text{PR} = 1.28837 A -0.71753 V + 0.07208 I_2 + 5.18969$ & 0.03548 & 3 & 0.98354 \\
$S_\text{1b}$ &$\hat{y}_\text{1b}^\text{PR} = 1.36089 A -0.61041 V + 0.15561 I + 5.31028$ & 0.01995 & 3 & 0.98697 \\
$S_\text{2a}$ & $\hat{y}_\text{2a}^\text{PR}  = 1.61574 A -19.35740 V + 0.06419 I + 101.58972$ & 0.97724 & 3 & 0.98743 \\
$S_\text{2b}$ & $\hat{y}_\text{2b}^\text{PR} =1.61494 A - 19.42096 V + 0.06494 I + 101.84952$ & 0.97724 & 3 & 0.98740 \\
\hline
\end{tabular}
}
\caption{
Optimal candidate formulas for the minimum volume for toric Calabi-Yau 3-folds given by $y=1/V_{min}$ and obtained under L1 (Lasso) regularized polynomial regression (PR) on datasets $S_\text{1a}$, $S_\text{1b}$, $S_\text{2a}$ and $S_\text{2b}$. For each optimal candidate formula, we give the optimal regularization parameter $\alpha^*$ that maximizes the corresponding $R^2$-score and minimizes the number of non-zero coefficients $N_{\beta_a}$ in the formula. 
}
\label{t_data02}
\end{table*}

\begin{table*}[ht!!]
\centering
\resizebox{0.86\hsize}{!}{
\begin{tabular}{|c|c|c|c|c|}
\hline
data set & $y=1/V_{min}$ & $\alpha^*$ & $N_{\beta_a(\alpha^*)}$ & $R^2(\alpha^*)$ 
\\
\hline \hline
$S_\text{1a}$ & $\hat{y}_\text{1a}^\text{LR} = 1.97348 A^{0.77011} V^{-0.21355} I_3^{0.08796} I_4^{0.02722} I_5^{0.00202} e^{0.00923 (\log{I_3})^2}$ & 0.00045 & 6 & 0.98932\\
$S_\text{1b}$ & $\hat{y}_\text{1b}^\text{LR} = 1.75668 A^{0.74154} V^{-0.182009} E^{0.00050} I_3^{0.16451} I_4^{0.00679} e^{0.00447 (\log {I_3})^2}$ & 0.00032 & 6 & 0.98992\\
$S_\text{2a}$ & $\hat{y}_\text{2a}^\text{LR} = 2.50772 A^{0.95411} V^{-0.21992} I_3^{0.02867}$ & 0.00112 & 3 & 0.99281 \\
$S_\text{2b}$ &  $\hat{y}_\text{2b}^\text{LR} = 2.51288 A^{0.95322} V^{-0.21970} I_3^{0.02898}$ & 0.00112 & 3 & 0.99297 \\
\hline
\end{tabular}
}
\caption{
Optimal candidate formulas for the minimum volume for toric Calabi-Yau 3-folds given by $y=1/V_{min}$ and obtained under L1 (Lasso) regularized logarithmic regression (LR) on datasets $S_\text{1a}$, $S_\text{1b}$, $S_\text{2a}$ and $S_\text{2b}$. For each optimal candidate formula, we give the optimal regularization parameter $\alpha^*$ that maximizes the corresponding $R^2$-score and minimizes the number of non-zero coefficients $N_{\beta_a}$ in the formula. 
}
\label{t_data03}
\end{table*}

\noindent
\textit{Parameter Sparsity vs Accuracy.}
Like in all regression problems, we introduce as a measure of how well the model fits the observed data using the $R^2$-score \cite{montgomery2021introduction,hastie2009elements} given by, 
\beal{es6a01}
R^2 = 1 - \frac{S_{res}}{S_{tot}} ~,~
\eea
where the residual sum of squares $S_{res}$ is given by,
\beal{es6a02}
S_{res} = \sum_{j=1}^{N} (y^j - \hat{y}^j)^2
\eea
and the total sum of squares $S_{tot}$ is given by,
\beal{es6a03}
S_{tot} = \sum_{j=1}^{N} (y^j - \overline{y})^2~.~
\eea
Here, $\hat{y}^j$ denotes the predicted value for the minimum volume $V_{min}^j$ given by $y^j=1/V_{min}^j$, whereas $\overline{y}$ denotes the mean of the expected values $y^j$.

We recall that the optimization problem for the L1-regularized regression model is to minimize the loss function $\mathcal{L}+\Delta \mathcal{L}_{\text{L1}}$ with the L1 regularization term.
As we discussed in the sections above, this optimization problem focuses on minimizing the mean squared error with a penalty for non-zero coefficients $\beta_a(\alpha)$, which depends on the regularization parameter $\alpha$.

Here, we note that there is an additional optimization problem regarding the maximization of the $R^2$-score in \eref{es6a01} and the minimization of the number $N_{\beta_a(\alpha)}$ of non-zero coefficients $\beta_a(\alpha)$.
We can formulate this additional optimization problem as follows, 
\beal{es6a10}
\max_{\alpha} \left\{
R^2(\alpha) - \lambda \frac{N_{\beta_a(\alpha)}}{N_x}
\right\} ~,~
\eea
where $0 < N_{\beta_a(\alpha)} \leq N_x$, and the values of the coefficients $\beta_a(\alpha)$ and the $R^2(\alpha)$-score all depend on the regularization parameter $\alpha$.
$\lambda$ is a positive hyperparameter that regulates how much we value sparsity of feature coefficients $\beta_a(\alpha)$ over the accuracy of the estimate given by $R^2(\alpha)$.
\\

\noindent
\textit{Candidate Formulas.}
The candidate formulas for the minimum volume for toric Calabi-Yau 3-folds 
are identified by an optimal regularization parameter $\alpha^*$ that maximizes the $R^2$-score of the candidate formula and minimizes the number of non-zero coefficients $N_{\beta_a(\alpha)}$ corresponding to features in the chosen regression model.
In order to identify the optimal regularization parameter $\alpha^*$ for the optimization problem in \eref{es6a10}, 
we search for $\alpha^*$ in a given fixed range for $\alpha$ as specified in \fref{f_fig07} and \fref{f_fig08}.
We do the search for the optimal regularization parameter $\alpha^*$ for all four datasets in \tref{t_data01} for both L1-regularized polynomial regression and L1-regularized logarithmic regression as discussed in sections \sref{sback3} and \sref{sback4}.
The chosen L1-regularized regression models are trained for a particular value of the regularization parameter $\alpha$ under a fixed randomly chosen 80\% training and 20\% testing data split, where the corresponding $R^2$-score depending on $\alpha$ is obtained from the testing data.

\begin{figure*}[ht!!]
\begin{center}
\resizebox{0.84\hsize}{!}{
\includegraphics[height=5cm]{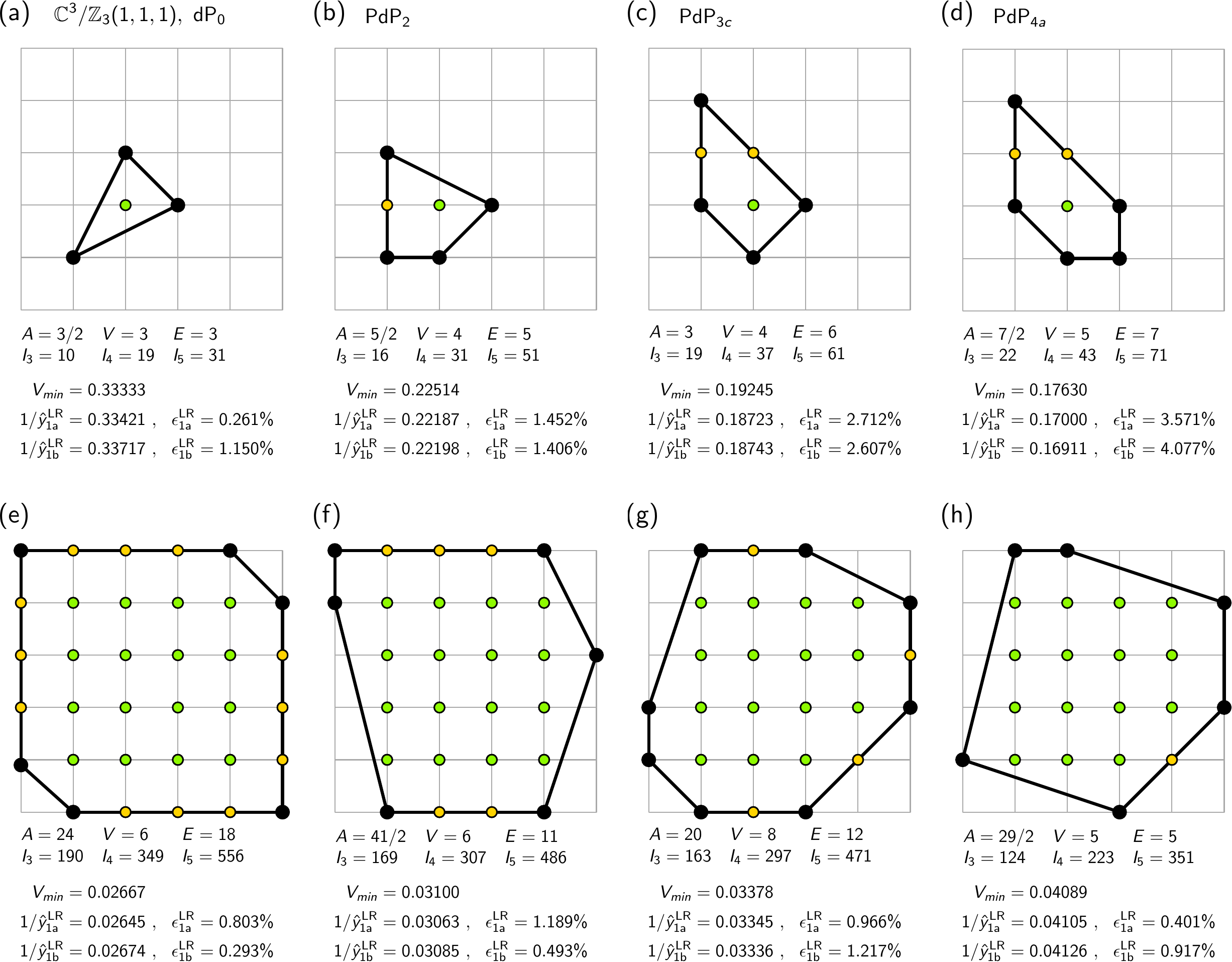} 
}
\caption{
The L1-regularized logarithmic regression models trained on datasets $S_\text{1a}$ and $S_\text{1b}$
perform better on toric diagrams with larger areas $A$ (see selection in (e)-(h)) than for toric diagrams with smaller areas $A$ (see selection in (a)-(d)). 
The performance is measure by the relative percentage error $\epsilon(1/\hat{y})$ of the predicted minimum volume given by $1/\hat{y}$.
The $R^2$-scores for the L1-regularized logarithmic regression models trained on datasets $S_\text{1a}$ and $S_\text{1b}$ are 
$R^2\left(y_\text{1a}^\text{LR}\right) = 0.98932$ and $R^2\left(y_\text{1b}^\text{LR}\right) = 0.98992$, respectively.
\label{f_fig09}}
 \end{center}
 \end{figure*}

\fref{f_fig07} shows respectively for datasets $S_\text{1a}$ and $S_\text{2a}$ plots for the L1 regularization parameter $\alpha$ for polynomial regression against standardized coefficients $\overline{\beta}_a(\alpha)$, against the number of non-zero coefficients $N_{\beta_a(\alpha)}$, and against the $R^2$-score.
Here, the standardized coefficients $\overline{\beta}_a(\alpha)$ are obtained when the training is conducted over normalized features $\overline{x}_a$.
When the training is completed for a specific value of $\alpha$, the candidate formula for the minimum volume given by $y=1/V_{min}$ is obtained by reversing the normalization on the features, giving us the coefficients $\beta_a(\alpha)$ of the candidate formula.
We also have \fref{f_fig08} which shows respectively for datasets $S_\text{1a}$ and $S_\text{2a}$ plots for the L1 regularization parameter $\alpha$ for logarithmic regression against the standardized coefficients $\overline{\beta}_a(\alpha)$, the number of non-zero coefficients $N_{\beta_a(\alpha)}$ and the $R^2$-score.
Similar plots can also be obtained for datasets $S_\text{1b}$ and $S_\text{2b}$ for both L1-regularized polynomial regression and L1-regularized logarithmic regression.

Overall, the plots illustrate that the identified optimal regularization parameters $\alpha^*$ minimize the number of non-zero coefficients $N_{\beta_a(\alpha)}$ in the formula estimating the minimum volume given by $y=1/V_{min}$, as well as maximize the accuracy of the formulas measured by the $R^2$-score.
\tref{t_data02} and \tref{t_data03} summarize respectively the most optimal candidate formulas for the minimum volume given by $y=1/V_{min}$ under L1-regularized polynomial regression and L1-regularized logarithmic regression for the four datasets in \tref{t_data01}, with the corresponding optimal regularization parameters $\alpha^*$, the corresponding number of non-zero coefficients $N_{\beta_a(\alpha)}$ and the $R^2$-score.

A closer look reveals that for all models, the identified optimal regularization parameters $\alpha^*$ results in formulas that approximate the minimum volume $y=1/V_{min}$ extremely well for all the datasets $S_\text{1a}$, $S_\text{1b}$, $S_\text{2a}$ and $S_\text{2b}$.
Overall, the L1-regularized logarithmic regression models seem to give more accurate results than the L1-regularized polynomial regression models with $N_{\beta_a(\alpha)} \leq 6$ over all datasets.
In particular, L1-regularized logarithmic regression models trained on datasets $S_\text{2a}$ and $S_\text{2b}$ have $R^2$-scores above $0.99$, which is exceptionally high.

Having a closer look at explicit examples of toric Calabi-Yau 3-folds in the datasets reveals however that the performances of the regularized regression models can vary between different toric Calabi-Yau 3-folds.
For example, focusing on the L1-regularized logarithmic regression models trained on $S_\text{1a}$ and $S_\text{1b}$, 
we observe that the minimum volumes given by $1/\hat{y}_\text{1a}^\text{LR}$ and $1/\hat{y}_\text{1b}^\text{LR}$ in \tref{t_data03} perform differently for toric diagrams with smaller areas $A$ compared to toric diagrams with larger areas $A$ as illustrated in \fref{f_fig09}.
Similar observations can be made for the L1-regularized logarithmic regression models trained on $S_\text{2a}$ and $S_\text{2b}$ as well as the L1-regularized polynomial regression models.

In summary, we can calculate the expected relative percentage errors $E[\epsilon]$ of the predicted minimum volumes given by $1/\hat{y}$ and the corresponding standard deviations $\sigma[\epsilon]$ for the L1-regularized logarithmic regression models as follows, 
\beal{es6a41}
E\left[\epsilon_\text{1a}^\text{LR}\right] = 2.158 \% ~,~
\sigma\left[\epsilon_\text{1a}^\text{LR}\right] = 1.696 \% ~,~
\nn\\
E\left[\epsilon_\text{1b}^\text{LR}\right] = 1.884 \% ~,~
\sigma\left[\epsilon_\text{1b}^\text{LR}\right] = 1.545 \% ~,~
\nn\\
E\left[\epsilon_\text{2a}^\text{LR}\right] = 3.577 \% ~,~
\sigma\left[\epsilon_\text{2a}^\text{LR}\right] = 2.396 \% ~,~
\nn\\
E\left[\epsilon_\text{2b}^\text{LR}\right] = 3.579 \% ~,~
\sigma\left[\epsilon_\text{2b}^\text{LR}\right] = 2.399 \% ~.~
\eea
We note that the models trained on $S_\text{2a}$ and $S_\text{2b}$ have a larger expected relative percentage error than the ones trained on $S_\text{1a}$ and $S_\text{1b}$.
This is partly due to the fact that $S_\text{2a}$ and $S_\text{2b}$ contain randomly selected toric diagrams in a $30\times 30$ lattice box in $\mathbb{Z}^2$ and $r=15$ circle, respectively, whereas $S_\text{1a}$ and $S_\text{1b}$ contain the full set of toric diagrams in a $5\times 5$ lattice box in $\mathbb{Z}^2$ and $r=3.5$ circle, respectively, as defined in \tref{t_data01}.

We also note that the $R^2$-scores of the L1-regularized logarithmic regression models in \tref{t_data03}, 
\beal{es6a42}
R^2\left(y_\text{1a}^\text{LR}\right) = 0.98932~,~
R^2\left(y_\text{1b}^\text{LR}\right) = 0.98992~,~\nn\\
R^2\left(y_\text{2a}^\text{LR}\right) = 0.99281~,~
R^2\left(y_\text{2b}^\text{LR}\right) = 0.99297~,~
\eea
are overall very high and close to $1$.
Compared to the expected relative percentage errors in \eref{es6a41}, which measure how far off predictions of the minimum volume given by $1/\hat{y}$ are, the $R^2$-score is a measure of the accuracy of the trained regression model. 
It quantifies the proportion of the variation in $y=1/V_{min}$ that can be predicted using the features selected from the corresponding toric diagrams of the toric Calabi-Yau 3-folds.
\\

\section{Discussions and Conclusions}\label{sback6}

With this work, we demonstrated that employing regularization in machine learning models can effectively address the limitations posed by supervised machine learning techniques applied to problems that occur in the context of string theory. 
In particular, we have shown that the minimum volume $V_{min}$ for Sasaki-Einstein 5-manifolds corresponding to toric Calabi-Yau 3-folds can be expressed 
by just 3 features of the associated toric diagrams $\Delta$ with an $R^2$-score $\geq 0.98$. 
These 3 features are the area $A$ of $\Delta$, the number of vertices $V$ in $\Delta$, and the number of internal points in the factor $n=3$ enlarged toric diagram $\Delta_3$.

The simultaneous maximization of the $R^2$-score and the minimization of the number surviving parameters in the candidate function for $y=1/V_{min}$
by varying the regularization strength given by the regularization parameter $\alpha$,
the proposed regularized regression models in this work 
give far more presentable, interpretable and explainable results than our previous work in \cite{Krefl:2017yox}.
Above all, as suggested in \fref{f_fig09}, the candidate formulas for the minimum volumes of toric Calabi-Yau 3-folds obtained in this study are concise enough to facilitate the examination of why some toric Calabi-Yau 3-folds are associated with minimum volumes that are more challenging to predict than those of certain other toric Calabi-Yau 3-folds.
We plan to report on these investigations in the near future. 
We foresee that the application of regularization schemes to other supervised machine learning applications in string theory will open up equally promising research opportunities in the future.

\acknowledgments

R.K.-S. would like to thank the Simons Center for Geometry and Physics at Stony Brook University,
the City University of New York Graduate Center,  
the Institute for Basic Science Center for Geometry and Physics,
as well as the Kavli Institute for the Physics and Mathematics of the Universe
for hospitality during various stages of this work.
He is supported by a Basic Research Grant of the National Research Foundation of Korea (NRF-2022R1F1A1073128).
He is also supported by a Start-up Research Grant for new faculty at UNIST (1.210139.01), a UNIST AI Incubator Grant (1.230038.01) and UNIST UBSI Grants (1.230168.01, 1.230078.01), as well as an Industry Research Project (2.220916.01) funded by Samsung SDS in Korea.  
He is also partly supported by the BK21 Program (``Next Generation Education Program for Mathematical Sciences'', 4299990414089) funded by the Ministry of Education in Korea and the National Research Foundation of Korea (NRF).


\bibliographystyle{jhep}
\bibliography{mybib}

\providecommand{\href}[2]{#2}\begingroup\raggedright\begin{thebibliography}{10}

\bibitem{He:2017aed}
Y.-H. He, \emph{{Deep-Learning the Landscape}},
  \href{http://arxiv.org/abs/1706.02714}{{\tt 1706.02714}}.

\bibitem{Krefl:2017yox}
D.~Krefl and R.-K. Seong, \emph{{Machine Learning of Calabi-Yau Volumes}},
  \href{http://dx.doi.org/10.1103/PhysRevD.96.066014}{\emph{Phys. Rev. D} {\bf
  96} (2017) 066014}, [\href{http://arxiv.org/abs/1706.03346}{{\tt
  1706.03346}}].

\bibitem{Ruehle:2017mzq}
F.~Ruehle, \emph{{Evolving neural networks with genetic algorithms to study the
  String Landscape}},
  \href{http://dx.doi.org/10.1007/JHEP08(2017)038}{\emph{JHEP} {\bf 08} (2017)
  038}, [\href{http://arxiv.org/abs/1706.07024}{{\tt 1706.07024}}].

\bibitem{Carifio:2017bov}
J.~Carifio, J.~Halverson, D.~Krioukov and B.~D. Nelson, \emph{{Machine Learning
  in the String Landscape}},
  \href{http://dx.doi.org/10.1007/JHEP09(2017)157}{\emph{JHEP} {\bf 09} (2017)
  157}, [\href{http://arxiv.org/abs/1707.00655}{{\tt 1707.00655}}].

\bibitem{Cole:2019enn}
A.~Cole, A.~Schachner and G.~Shiu, \emph{{Searching the Landscape of Flux Vacua
  with Genetic Algorithms}},
  \href{http://dx.doi.org/10.1007/JHEP11(2019)045}{\emph{JHEP} {\bf 11} (2019)
  045}, [\href{http://arxiv.org/abs/1907.10072}{{\tt 1907.10072}}].

\bibitem{Cole:2020gkd}
A.~Cole, G.~J. Loges and G.~Shiu, \emph{{Interpretable Phase Detection and
  Classification with Persistent Homology}},  in \emph{{34th Conference on
  Neural Information Processing Systems}}, 12, 2020.
\newblock \href{http://arxiv.org/abs/2012.00783}{{\tt 2012.00783}}.

\bibitem{Halverson:2020trp}
J.~Halverson, A.~Maiti and K.~Stoner, \emph{{Neural Networks and Quantum Field
  Theory}}, \href{http://dx.doi.org/10.1088/2632-2153/abeca3}{\emph{Mach.
  Learn. Sci. Tech.} {\bf 2} (2021) 035002},
  [\href{http://arxiv.org/abs/2008.08601}{{\tt 2008.08601}}].

\bibitem{Gukov:2020qaj}
S.~Gukov, J.~Halverson, F.~Ruehle and P.~Su\l{}kowski, \emph{{Learning to
  Unknot}}, \href{http://dx.doi.org/10.1088/2632-2153/abe91f}{\emph{Mach.
  Learn. Sci. Tech.} {\bf 2} (2021) 025035},
  [\href{http://arxiv.org/abs/2010.16263}{{\tt 2010.16263}}].

\bibitem{Abel:2021rrj}
S.~Abel, A.~Constantin, T.~R. Harvey and A.~Lukas, \emph{{Evolving Heterotic
  Gauge Backgrounds: Genetic Algorithms versus Reinforcement Learning}},
  \href{http://dx.doi.org/10.1002/prop.202200034}{\emph{Fortsch. Phys.} {\bf
  70} (2022) 2200034}, [\href{http://arxiv.org/abs/2110.14029}{{\tt
  2110.14029}}].

\bibitem{Krippendorf:2021uxu}
S.~Krippendorf, R.~Kroepsch and M.~Syvaeri, \emph{{Revealing systematics in
  phenomenologically viable flux vacua with reinforcement learning}},
  \href{http://arxiv.org/abs/2107.04039}{{\tt 2107.04039}}.

\bibitem{Cole:2021nnt}
A.~Cole, S.~Krippendorf, A.~Schachner and G.~Shiu, \emph{{Probing the Structure
  of String Theory Vacua with Genetic Algorithms and Reinforcement Learning}},
  in \emph{{35th Conference on Neural Information Processing Systems}}, 11,
  2021.
\newblock \href{http://arxiv.org/abs/2111.11466}{{\tt 2111.11466}}.

\bibitem{Berglund:2023ztk}
P.~Berglund, Y.-H. He, E.~Heyes, E.~Hirst, V.~Jejjala and A.~Lukas, \emph{{New
  Calabi-Yau Manifolds from Genetic Algorithms}},
  \href{http://arxiv.org/abs/2306.06159}{{\tt 2306.06159}}.

\bibitem{Demirtas:2023fir}
M.~Demirtas, J.~Halverson, A.~Maiti, M.~D. Schwartz and K.~Stoner,
  \emph{{Neural Network Field Theories: Non-Gaussianity, Actions, and
  Locality}},  \href{http://arxiv.org/abs/2307.03223}{{\tt 2307.03223}}.

\bibitem{Bull:2018uow}
K.~Bull, Y.-H. He, V.~Jejjala and C.~Mishra, \emph{{Machine Learning CICY
  Threefolds}},
  \href{http://dx.doi.org/10.1016/j.physletb.2018.08.008}{\emph{Phys. Lett. B}
  {\bf 785} (2018) 65--72}, [\href{http://arxiv.org/abs/1806.03121}{{\tt
  1806.03121}}].

\bibitem{Jejjala:2019kio}
V.~Jejjala, A.~Kar and O.~Parrikar, \emph{{Deep Learning the Hyperbolic Volume
  of a Knot}},
  \href{http://dx.doi.org/10.1016/j.physletb.2019.135033}{\emph{Phys. Lett. B}
  {\bf 799} (2019) 135033}, [\href{http://arxiv.org/abs/1902.05547}{{\tt
  1902.05547}}].

\bibitem{Brodie:2019dfx}
C.~R. Brodie, A.~Constantin, R.~Deen and A.~Lukas, \emph{{Machine Learning Line
  Bundle Cohomology}},
  \href{http://dx.doi.org/10.1002/prop.201900087}{\emph{Fortsch. Phys.} {\bf
  68} (2020) 1900087}, [\href{http://arxiv.org/abs/1906.08730}{{\tt
  1906.08730}}].

\bibitem{He:2020lbz}
Y.-H. He and A.~Lukas, \emph{{Machine Learning Calabi-Yau Four-folds}},
  \href{http://dx.doi.org/10.1016/j.physletb.2021.136139}{\emph{Phys. Lett. B}
  {\bf 815} (2021) 136139}, [\href{http://arxiv.org/abs/2009.02544}{{\tt
  2009.02544}}].

\bibitem{Erbin:2020tks}
H.~Erbin and R.~Finotello, \emph{{Machine learning for complete intersection
  Calabi-Yau manifolds: a methodological study}},
  \href{http://dx.doi.org/10.1103/PhysRevD.103.126014}{\emph{Phys. Rev. D} {\bf
  103} (2021) 126014}, [\href{http://arxiv.org/abs/2007.15706}{{\tt
  2007.15706}}].

\bibitem{Anagiannis:2021cco}
V.~Anagiannis and M.~C.~N. Cheng, \emph{{Entangled q-convolutional neural
  nets}}, \href{http://dx.doi.org/10.1088/2632-2153/ac2800}{\emph{Mach. Learn.
  Sci. Tech.} {\bf 2} (2021) 045026},
  [\href{http://arxiv.org/abs/2103.11785}{{\tt 2103.11785}}].

\bibitem{Larfors:2022nep}
M.~Larfors, A.~Lukas, F.~Ruehle and R.~Schneider, \emph{{Numerical metrics for
  complete intersection and Kreuzer\textendash{}Skarke Calabi\textendash{}Yau
  manifolds}}, \href{http://dx.doi.org/10.1088/2632-2153/ac8e4e}{\emph{Mach.
  Learn. Sci. Tech.} {\bf 3} (2022) 035014},
  [\href{http://arxiv.org/abs/2205.13408}{{\tt 2205.13408}}].

\bibitem{Krippendorf:2020gny}
S.~Krippendorf and M.~Syvaeri, \emph{{Detecting Symmetries with Neural
  Networks}},  \href{http://arxiv.org/abs/2003.13679}{{\tt 2003.13679}}.

\bibitem{Berman:2021mcw}
D.~S. Berman, Y.-H. He and E.~Hirst, \emph{{Machine learning Calabi-Yau
  hypersurfaces}},
  \href{http://dx.doi.org/10.1103/PhysRevD.105.066002}{\emph{Phys. Rev. D} {\bf
  105} (2022) 066002}, [\href{http://arxiv.org/abs/2112.06350}{{\tt
  2112.06350}}].

\bibitem{Bao:2021olg}
J.~Bao, Y.-H. He and E.~Hirst, \emph{{Neurons on Amoebae}},
  \href{http://dx.doi.org/10.1016/j.jsc.2022.08.021}{\emph{J. Symb. Comput.}
  {\bf 116} (2022) 1--38}, [\href{http://arxiv.org/abs/2106.03695}{{\tt
  2106.03695}}].

\bibitem{Seong:2023njx}
R.-K. Seong, \emph{{Unsupervised Machine Learning Techniques for Exploring
  Tropical Coamoeba, Brane Tilings and Seiberg Duality}},
  \href{http://arxiv.org/abs/2309.05702}{{\tt 2309.05702}}.

\bibitem{Martelli:2006yb}
D.~Martelli, J.~Sparks and S.-T. Yau, \emph{{Sasaki-Einstein manifolds and
  volume minimisation}},
  \href{http://dx.doi.org/10.1007/s00220-008-0479-4}{\emph{Commun. Math. Phys.}
  {\bf 280} (2008) 611--673}, [\href{http://arxiv.org/abs/hep-th/0603021}{{\tt
  hep-th/0603021}}].

\bibitem{Martelli:2005tp}
D.~Martelli, J.~Sparks and S.-T. Yau, \emph{{The geometric dual of
  a-maximisation for toric Sasaki- Einstein manifolds}},
  \href{http://dx.doi.org/10.1007/s00220-006-0087-0}{\emph{Commun. Math. Phys.}
  {\bf 268} (2006) 39--65}, [\href{http://arxiv.org/abs/hep-th/0503183}{{\tt
  hep-th/0503183}}].

\bibitem{fulton}
W.~Fulton, \emph{{Introduction to toric varieties}}.
\newblock Annals of mathematics studies. Princeton Univ. Press, Princeton, NJ,
  1993.

\bibitem{1997hep.th...11013L}
N.~C. {Leung} and C.~{Vafa}, \emph{{Branes and Toric Geometry}}, {\emph{ArXiv
  High Energy Physics - Theory e-prints} (Nov., 1997) },
  [\href{http://arxiv.org/abs/hep-th/9711013}{{\tt hep-th/9711013}}].

\bibitem{Greene:1996cy}
B.~R. Greene, \emph{{String theory on Calabi-Yau manifolds}},  in
  \emph{{Theoretical Advanced Study Institute in Elementary Particle Physics
  (TASI 96): Fields, Strings, and Duality}}, pp.~543--726, 6, 1996.
\newblock \href{http://arxiv.org/abs/hep-th/9702155}{{\tt hep-th/9702155}}.

\bibitem{Douglas:1997de}
M.~R. Douglas, B.~R. Greene and D.~R. Morrison, \emph{{Orbifold resolution by
  D-branes}},
  \href{http://dx.doi.org/10.1016/S0550-3213(97)00517-8}{\emph{Nucl.Phys.} {\bf
  B506} (1997) 84--106}, [\href{http://arxiv.org/abs/hep-th/9704151}{{\tt
  hep-th/9704151}}].

\bibitem{Witten:1998qj}
E.~Witten, \emph{{Anti-de Sitter space and holography}},
  \href{http://dx.doi.org/10.4310/ATMP.1998.v2.n2.a2}{\emph{Adv. Theor. Math.
  Phys.} {\bf 2} (1998) 253--291},
  [\href{http://arxiv.org/abs/hep-th/9802150}{{\tt hep-th/9802150}}].

\bibitem{Klebanov:1998hh}
I.~R. Klebanov and E.~Witten, \emph{{Superconformal field theory on
  three-branes at a Calabi-Yau singularity}},
  \href{http://dx.doi.org/10.1016/S0550-3213(98)00654-3}{\emph{Nucl.Phys.} {\bf
  B536} (1998) 199--218}, [\href{http://arxiv.org/abs/hep-th/9807080}{{\tt
  hep-th/9807080}}].

\bibitem{Douglas:1996sw}
M.~R. Douglas and G.~W. Moore, \emph{{D-branes, Quivers, and ALE Instantons}},
  \href{http://arxiv.org/abs/hep-th/9603167}{{\tt hep-th/9603167}}.

\bibitem{Lawrence:1998ja}
A.~E. Lawrence, N.~Nekrasov and C.~Vafa, \emph{{On conformal field theories in
  four-dimensions}},
  \href{http://dx.doi.org/10.1016/S0550-3213(98)00495-7}{\emph{Nucl.Phys.} {\bf
  B533} (1998) 199--209}, [\href{http://arxiv.org/abs/hep-th/9803015}{{\tt
  hep-th/9803015}}].

\bibitem{Feng:2000mi}
B.~Feng, A.~Hanany and Y.-H. He, \emph{{D-brane gauge theories from toric
  singularities and toric duality}},
  \href{http://dx.doi.org/10.1016/S0550-3213(00)00699-4}{\emph{Nucl. Phys.}
  {\bf B595} (2001) 165--200}, [\href{http://arxiv.org/abs/hep-th/0003085}{{\tt
  hep-th/0003085}}].

\bibitem{Feng:2001xr}
B.~Feng, A.~Hanany and Y.-H. He, \emph{{Phase structure of D-brane gauge
  theories and toric duality}},
  \href{http://dx.doi.org/10.1088/1126-6708/2001/08/040}{\emph{JHEP} {\bf 08}
  (2001) 040}, [\href{http://arxiv.org/abs/hep-th/0104259}{{\tt
  hep-th/0104259}}].

\bibitem{Maldacena:1997re}
J.~M. Maldacena, \emph{{The large N limit of superconformal field theories and
  supergravity}}, \href{http://dx.doi.org/10.1023/A:1026654312961}{\emph{Adv.
  Theor. Math. Phys.} {\bf 2} (1998) 231--252},
  [\href{http://arxiv.org/abs/hep-th/9711200}{{\tt hep-th/9711200}}].

\bibitem{Morrison:1998cs}
D.~R. Morrison and M.~R. Plesser, \emph{{Nonspherical horizons. 1.}},
  {\emph{Adv.Theor.Math.Phys.} {\bf 3} (1999) 1--81},
  [\href{http://arxiv.org/abs/hep-th/9810201}{{\tt hep-th/9810201}}].

\bibitem{Acharya:1998db}
B.~S. Acharya, J.~M. Figueroa-O'Farrill, C.~M. Hull and B.~J. Spence,
  \emph{{Branes at conical singularities and holography}}, {\emph{Adv. Theor.
  Math. Phys.} {\bf 2} (1999) 1249--1286},
  [\href{http://arxiv.org/abs/hep-th/9808014}{{\tt hep-th/9808014}}].

\bibitem{Intriligator:2003jj}
K.~A. Intriligator and B.~Wecht, \emph{{The Exact superconformal R symmetry
  maximizes a}},
  \href{http://dx.doi.org/10.1016/S0550-3213(03)00459-0}{\emph{Nucl. Phys. B}
  {\bf 667} (2003) 183--200}, [\href{http://arxiv.org/abs/hep-th/0304128}{{\tt
  hep-th/0304128}}].

\bibitem{Butti:2005vn}
A.~Butti and A.~Zaffaroni, \emph{{R-charges from toric diagrams and the
  equivalence of a- maximization and Z-minimization}},
  \href{http://dx.doi.org/10.1088/1126-6708/2005/11/019}{\emph{JHEP} {\bf 11}
  (2005) 019}, [\href{http://arxiv.org/abs/hep-th/0506232}{{\tt
  hep-th/0506232}}].

\bibitem{Butti:2005ps}
A.~Butti and A.~Zaffaroni, \emph{{From toric geometry to quiver gauge theory:
  The Equivalence of a-maximization and Z-minimization}},
  \href{http://dx.doi.org/10.1002/prop.200510276}{\emph{Fortsch.Phys.} {\bf 54}
  (2006) 309--316}, [\href{http://arxiv.org/abs/hep-th/0512240}{{\tt
  hep-th/0512240}}].

\bibitem{Gubser:1998vd}
S.~S. Gubser, \emph{{Einstein manifolds and conformal field theories}},
  \href{http://dx.doi.org/10.1103/PhysRevD.59.025006}{\emph{Phys. Rev. D} {\bf
  59} (1999) 025006}, [\href{http://arxiv.org/abs/hep-th/9807164}{{\tt
  hep-th/9807164}}].

\bibitem{Henningson:1998gx}
M.~Henningson and K.~Skenderis, \emph{{The Holographic Weyl anomaly}},
  \href{http://dx.doi.org/10.1088/1126-6708/1998/07/023}{\emph{JHEP} {\bf 07}
  (1998) 023}, [\href{http://arxiv.org/abs/hep-th/9806087}{{\tt
  hep-th/9806087}}].

\bibitem{Benvenuti:2006qr}
S.~Benvenuti, B.~Feng, A.~Hanany and Y.-H. He, \emph{{Counting BPS operators in
  gauge theories: Quivers, syzygies and plethystics}},
  \href{http://dx.doi.org/10.1088/1126-6708/2007/11/050}{\emph{JHEP} {\bf 11}
  (2007) 050}, [\href{http://arxiv.org/abs/hep-th/0608050}{{\tt
  hep-th/0608050}}].

\bibitem{Feng:2007ur}
B.~Feng, A.~Hanany and Y.-H. He, \emph{{Counting Gauge Invariants: the
  Plethystic Program}},
  \href{http://dx.doi.org/10.1088/1126-6708/2007/03/090}{\emph{JHEP} {\bf 03}
  (2007) 090}, [\href{http://arxiv.org/abs/hep-th/0701063}{{\tt
  hep-th/0701063}}].

\bibitem{gauss1823theoria}
C.-F. Gauss, \emph{Theoria combinationis observationum erroribus minimis
  obnoxiae}.
\newblock Henricus Dieterich, 1823.

\bibitem{fisher1922mathematical}
R.~A. Fisher, \emph{On the mathematical foundations of theoretical statistics},
  {\emph{Philosophical transactions of the Royal Society of London. Series A,
  containing papers of a mathematical or physical character} {\bf 222} (1922)
  309--368}.

\bibitem{mendenhall2003second}
W.~Mendenhall, T.~Sincich and N.~S. Boudreau, \emph{A second course in
  statistics: regression analysis}, vol.~6.
\newblock Prentice Hall Upper Saddle River, NJ, 2003.

\bibitem{freedman2009statistical}
D.~A. Freedman, \emph{Statistical models: theory and practice}.
\newblock cambridge university press, 2009.

\bibitem{jobson2012applied}
J.~D. Jobson, \emph{Applied multivariate data analysis: regression and
  experimental design}.
\newblock Springer Science \& Business Media, 2012.

\bibitem{lecun1998gradient}
Y.~LeCun, L.~Bottou, Y.~Bengio and P.~Haffner, \emph{Gradient-based learning
  applied to document recognition}, {\emph{Proceedings of the IEEE} {\bf 86}
  (1998) 2278--2324}.

\bibitem{krizhevsky2012imagenet}
A.~Krizhevsky, I.~Sutskever and G.~E. Hinton, \emph{Imagenet classification
  with deep convolutional neural networks}, {\emph{Advances in neural
  information processing systems} {\bf 25} (2012) }.

\bibitem{lecun2015deep}
Y.~LeCun, Y.~Bengio and G.~Hinton, \emph{Deep learning}, {\emph{nature} {\bf
  521} (2015) 436--444}.

\bibitem{schmidhuber2015deep}
J.~Schmidhuber, \emph{Deep learning in neural networks: An overview},
  {\emph{Neural networks} {\bf 61} (2015) 85--117}.

\bibitem{rumelhart1986learning}
D.~E. Rumelhart, G.~E. Hinton and R.~J. Williams, \emph{Learning
  representations by back-propagating errors}, {\emph{nature} {\bf 323} (1986)
  533--536}.

\bibitem{hastie2009elements}
T.~Hastie, R.~Tibshirani, J.~H. Friedman and J.~H. Friedman, \emph{The elements
  of statistical learning: data mining, inference, and prediction}, vol.~2.
\newblock Springer, 2009.

\bibitem{Hori:2000kt}
K.~Hori and C.~Vafa, \emph{{Mirror symmetry}},
  \href{http://arxiv.org/abs/hep-th/0002222}{{\tt hep-th/0002222}}.

\bibitem{Feng:2005gw}
B.~Feng, Y.-H. He, K.~D. Kennaway and C.~Vafa, \emph{{Dimer models from mirror
  symmetry and quivering amoebae}},
  \href{http://dx.doi.org/10.4310/ATMP.2008.v12.n3.a2}{\emph{Adv. Theor. Math.
  Phys.} {\bf 12} (2008) 489--545},
  [\href{http://arxiv.org/abs/hep-th/0511287}{{\tt hep-th/0511287}}].

\bibitem{tikhonov1963regularization}
A.~Tikhonov, \emph{Regularization of incorrectly posed problems},  in
  \emph{Soviet Math. Dokl.}, pp.~1624--1627, 1963.

\bibitem{tibshirani1996regression}
R.~Tibshirani, \emph{Regression shrinkage and selection via the lasso},
  {\emph{Journal of the Royal Statistical Society Series B: Statistical
  Methodology} {\bf 58} (1996) 267--288}.

\bibitem{Martelli:2004wu}
D.~Martelli and J.~Sparks, \emph{{Toric geometry, Sasaki-Einstein manifolds and
  a new infinite class of AdS/CFT duals}},
  \href{http://dx.doi.org/10.1007/s00220-005-1425-3}{\emph{Commun. Math. Phys.}
  {\bf 262} (2006) 51--89}, [\href{http://arxiv.org/abs/hep-th/0411238}{{\tt
  hep-th/0411238}}].

\bibitem{Benvenuti:2004dy}
S.~Benvenuti, S.~Franco, A.~Hanany, D.~Martelli and J.~Sparks, \emph{{An
  infinite family of superconformal quiver gauge theories with Sasaki-Einstein
  duals}}, \href{http://dx.doi.org/10.1088/1126-6708/2005/06/064}{\emph{JHEP}
  {\bf 06} (2005) 064}, [\href{http://arxiv.org/abs/hep-th/0411264}{{\tt
  hep-th/0411264}}].

\bibitem{Benvenuti:2005ja}
S.~Benvenuti and M.~Kruczenski, \emph{{From Sasaki-Einstein spaces to quivers
  via BPS geodesics: L**p,q|r}},
  \href{http://dx.doi.org/10.1088/1126-6708/2006/04/033}{\emph{JHEP} {\bf 04}
  (2006) 033}, [\href{http://arxiv.org/abs/hep-th/0505206}{{\tt
  hep-th/0505206}}].

\bibitem{Butti:2005sw}
A.~Butti, D.~Forcella and A.~Zaffaroni, \emph{{The Dual superconformal theory
  for L**pqr manifolds}},
  \href{http://dx.doi.org/10.1088/1126-6708/2005/09/018}{\emph{JHEP} {\bf 09}
  (2005) 018}, [\href{http://arxiv.org/abs/hep-th/0505220}{{\tt
  hep-th/0505220}}].

\bibitem{Franco:2005rj}
S.~Franco, A.~Hanany, K.~D. Kennaway, D.~Vegh and B.~Wecht, \emph{{Brane Dimers
  and Quiver Gauge Theories}},
  \href{http://dx.doi.org/10.1088/1126-6708/2006/01/096}{\emph{JHEP} {\bf 01}
  (2006) 096}, [\href{http://arxiv.org/abs/hep-th/0504110}{{\tt
  hep-th/0504110}}].

\bibitem{Hanany:2005ve}
A.~Hanany and K.~D. Kennaway, \emph{{Dimer models and toric diagrams}},
  \href{http://arxiv.org/abs/hep-th/0503149}{{\tt hep-th/0503149}}.

\bibitem{Franco:2005sm}
S.~Franco et~al., \emph{{Gauge theories from toric geometry and brane
  tilings}}, \href{http://dx.doi.org/10.1088/1126-6708/2006/01/128}{\emph{JHEP}
  {\bf 01} (2006) 128}, [\href{http://arxiv.org/abs/hep-th/0505211}{{\tt
  hep-th/0505211}}].

\bibitem{2003math.....10326K}
R.~{Kenyon}, \emph{{An introduction to the dimer model}}, {\emph{ArXiv
  Mathematics e-prints} (Oct., 2003) },
  [\href{http://arxiv.org/abs/math/0310326}{{\tt math/0310326}}].

\bibitem{kasteleyn1967graph}
P.~Kasteleyn, \emph{Graph theory and crystal physics}, {\emph{Graph theory and
  theoretical physics} (1967) 43--110}.

\bibitem{hirzebruch1968singularities}
F.~Hirzebruch, \emph{Singularities and exotic spheres}.
\newblock Societe Mathematic de France, 1968.

\bibitem{brieskorn1966beispiele}
E.~Brieskorn, \emph{Beispiele zur differentialtopologie von
  singularit{\"a}ten}, {\emph{Inventiones mathematicae} {\bf 2} (1966) 1--14}.

\bibitem{Witten:1993yc}
E.~Witten, \emph{{Phases of N = 2 theories in two dimensions}},
  \href{http://dx.doi.org/10.1016/0550-3213(93)90033-L}{\emph{Nucl. Phys.} {\bf
  B403} (1993) 159--222}, [\href{http://arxiv.org/abs/hep-th/9301042}{{\tt
  hep-th/9301042}}].

\bibitem{Butti:2007jv}
A.~Butti, D.~Forcella, A.~Hanany, D.~Vegh and A.~Zaffaroni, \emph{{Counting
  Chiral Operators in Quiver Gauge Theories}},
  \href{http://dx.doi.org/10.1088/1126-6708/2007/11/092}{\emph{JHEP} {\bf 0711}
  (2007) 092}, [\href{http://arxiv.org/abs/0705.2771}{{\tt 0705.2771}}].

\bibitem{Hanany:2010zz}
A.~Hanany and A.~Zaffaroni, \emph{{The master space of supersymmetric gauge
  theories}}, \href{http://dx.doi.org/10.1155/2010/427891}{\emph{Adv.High
  Energy Phys.} {\bf 2010} (2010) 427891}.

\bibitem{Forcella:2008bb}
D.~Forcella, A.~Hanany, Y.-H. He and A.~Zaffaroni, \emph{{The Master Space of
  N=1 Gauge Theories}},
  \href{http://dx.doi.org/10.1088/1126-6708/2008/08/012}{\emph{JHEP} {\bf 0808}
  (2008) 012}, [\href{http://arxiv.org/abs/0801.1585}{{\tt 0801.1585}}].

\bibitem{Forcella:2008eh}
D.~Forcella, A.~Hanany, Y.-H. He and A.~Zaffaroni, \emph{{Mastering the Master
  Space}},
  \href{http://dx.doi.org/10.1007/s11005-008-0255-6}{\emph{Lett.Math.Phys.}
  {\bf 85} (2008) 163--171}, [\href{http://arxiv.org/abs/0801.3477}{{\tt
  0801.3477}}].

\bibitem{Pouliot:1998yv}
P.~Pouliot, \emph{{Molien function for duality}},
  \href{http://dx.doi.org/10.1088/1126-6708/1999/01/021}{\emph{JHEP} {\bf 01}
  (1999) 021}, [\href{http://arxiv.org/abs/hep-th/9812015}{{\tt
  hep-th/9812015}}].

\bibitem{Seiberg:1994pq}
N.~Seiberg, \emph{{Electric - magnetic duality in supersymmetric nonAbelian
  gauge theories}},
  \href{http://dx.doi.org/10.1016/0550-3213(94)00023-8}{\emph{Nucl. Phys.} {\bf
  B435} (1995) 129--146}, [\href{http://arxiv.org/abs/hep-th/9411149}{{\tt
  hep-th/9411149}}].

\bibitem{2001JHEP...12..001B}
C.~E. {Beasley} and M.~{Ronen Plesser}, \emph{{Toric duality is Seiberg
  duality}},
  \href{http://dx.doi.org/10.1088/1126-6708/2001/12/001}{\emph{Journal of High
  Energy Physics} {\bf 12} (Dec., 2001) 1--+},
  [\href{http://arxiv.org/abs/hep-th/0109053}{{\tt hep-th/0109053}}].

\bibitem{goodfellow2016deep}
I.~Goodfellow, Y.~Bengio and A.~Courville, \emph{Deep learning}.
\newblock MIT press, 2016.

\bibitem{pick1899geometrisches}
G.~Pick, \emph{Geometrisches zur zahlenlehre}, {\emph{Sitzenber. Lotos
  (Prague)} {\bf 19} (1899) 311--319}.

\bibitem{Berglund:2021ztg}
P.~Berglund, B.~Campbell and V.~Jejjala, \emph{{Machine Learning Kreuzer-Skarke
  Calabi-Yau Threefolds}},  \href{http://arxiv.org/abs/2112.09117}{{\tt
  2112.09117}}.

\bibitem{montgomery2021introduction}
D.~C. Montgomery, E.~A. Peck and G.~G. Vining, \emph{Introduction to linear
  regression analysis}.
\newblock John Wiley \& Sons, 2021.

\bibitem{hoerl1970ridge}
A.~E. Hoerl and R.~W. Kennard, \emph{Ridge regression: Biased estimation for
  nonorthogonal problems}, {\emph{Technometrics} {\bf 12} (1970) 55--67}.

\bibitem{zou2005regularization}
H.~Zou and T.~Hastie, \emph{Regularization and variable selection via the
  elastic net}, {\emph{Journal of the Royal Statistical Society Series B:
  Statistical Methodology} {\bf 67} (2005) 301--320}.

\end{thebibliography}\endgroup

\end{document}